\documentclass[%
 aip,
 jmp,%
 amsmath,amssymb,
 reprint,%
]{revtex4-1}

\usepackage{graphicx}
\usepackage{dcolumn}
\usepackage{bm}

\usepackage{graphicx}
\usepackage{dcolumn}
\usepackage{bm}

\usepackage{lipsum}
\usepackage{color}
\usepackage[normalem]{ulem}
\usepackage{simplewick}
\usepackage{qcircuit}
\usepackage{braket}
\usepackage{float}
\usepackage{multirow}
\usepackage{tikz}
\usepackage[colorlinks=true,urlcolor=blue,citecolor=blue,linkcolor=blue]{hyperref}
\usepackage{amsmath}
\usepackage{amssymb}
\usepackage{tabularx}
\usepackage{soul}

\usepackage[ruled, vlined]{algorithm2e}

\usepackage{chemformula}
\usepackage[version=3]{mhchem} 

\begin{document}


\title{Quantum Orbital-Optimized Unitary Coupled Cluster Methods in the Strongly Correlated Regime: 
Can Quantum Algorithms Outperform their Classical Equivalents?}

\author{Igor O. Sokolov}
\email{iso@zurich.ibm.com}
\affiliation{IBM Research GmbH, Zurich Research Laboratory, S\"aumerstrasse 4, 8803 R\"uschlikon, Switzerland}

\author{Panagiotis Kl. Barkoutsos} 
\affiliation{IBM Research GmbH, Zurich Research Laboratory, S\"aumerstrasse 4, 8803 R\"uschlikon, Switzerland}

\author{Pauline J. Ollitrault}
\affiliation{IBM Research GmbH, Zurich Research Laboratory, S\"aumerstrasse 4, 8803 R\"uschlikon, Switzerland}
\affiliation{ETH Z\"urich, Laboratory of Physical Chemistry,
Vladimir-Prelog-Weg 2, 8093 Z\"urich, Switzerland}

\author{Donny Greenberg} 
\affiliation{IBM Thomas J. Watson Research Center, Yorktown Heights, New York 10598, USA}

\author{Julia Rice} 
\affiliation{IBM Almaden Research Center, San Jose, California 95120, USA}

\author{Marco Pistoia} 
\affiliation{IBM Thomas J. Watson Research Center, Yorktown Heights, New York 10598, USA}
\affiliation{JPMorgan Chase $\&$ Co, New York, New York 10172, USA}

\author{Ivano Tavernelli}
\email{ita@zurich.ibm.com}
\affiliation{IBM Research GmbH, Zurich Research Laboratory, S\"aumerstrasse 4, 8803 R\"uschlikon, Switzerland}

\date{\today}

\begin{abstract}

The Coupled Cluster (CC) method is used to compute the electronic correlation energy in atoms and molecules and often leads to highly accurate results.
However, due to its single-reference nature, standard CC in its projected form fails to describe quantum states characterized by strong electronic correlations and multi-reference projective methods become necessary.
On the other hand, quantum algorithms for the solution of many-electron problems have also emerged recently. 
The quantum UCC with singles and doubles (q-UCCSD) is a popular wavefunction Ansatz for the Variational Quantum Eigensolver (VQE) algorithm.
The variational nature of this approach can lead to significant advantages compared to its classical equivalent in the projected form, in particular for the description of strong electronic correlation.
However, due to the large number of gate operations required in q-UCCSD, approximations need to be introduced in order to make this approach implementable in a state-of-the-art quantum computer.
In this work, we evaluate several variants of the standard q-UCCSD Ansatz in which only a subset of excitations is included. 
In particular, we investigate the singlet and pair q-UCCD approaches combined with orbital optimization. 
We show that these approaches can capture the dissociation/distortion profiles of challenging systems such as \ce{H$_4$}, \ce{H$_2$O} and  \ce{N$_2$} molecules, as well as the one-dimensional periodic Fermi-Hubbard chain.
These results promote the future use of q-UCC methods for the solution of challenging electronic structure problems in quantum chemistry.

\end{abstract}

\maketitle

\section{\label{sec:intro}Introduction}

One of the main goals of computational quantum chemistry is the design of efficient methods for the calculation of the ground and excited state properties of molecular systems.
When combined with experimental results, first-principle (or \textit{ab-initio}) calculations enable the investigation of chemical and industrial processes 
(e.g. catalysis, electrochemistry, polymerization, and photochemistry to mention only a few) 
as well as the discovery of new materials and catalysts~\cite{Hachmann2010,Olivares-Amaya2011,Krein2012,Curtarolo2013,Ling2015, Reiher2016}. 

The ground state energy of a molecular or solid state system can be obtained from the solution of the corresponding Schr\"odinger equation. 
However, the solution space (i.e., the Hilbert space) grows exponentially with the system size, $N$, (e.g. the number of electrons and basis functions), making the exact solution of this problem intractable for systems with more than a few atoms~\cite{Helgaker2014}. 
Relying on a mean field approach, the Hartree-Fock (HF) method allows to efficiently compute ($\mathcal{O}(N^4)$) an approximation of the ground state which does not include any electronic correlation effects~\cite{Szabo1982}.
The correction to the energy can then be computed with the so-called post-HF methods. 
Among the most popular ones, we find M{\o}ller-Plesset (MP) perturbation theory~\cite{Moller1934, Ayala1999,Leininger2000,Grimme2000}, Configuration Interaction (CI)~\cite{Sherrill1999} and Coupled Cluster (CC)~\cite{vcivzek1966correlation, kummel2002biography, Bartlett2007}.
In particular, in CC the exponential Ansatz allows the systematic introduction of higher order configurations and makes the approach size extensive.
The CC method including single, double and an approximate treatment of the triple excitations, named CCSD(T)~\cite{Raghavachari1989}, scales as $\mathcal{O}(N^7)$~\cite{Bartlett2007} and is often regarded as the \textit{gold standard} for quantum chemistry calculations. 
Usually the projective CC equations~\cite{vcivzek1966correlation} are iteratively solved using, for instance, the quasi-Newton and direct inversion iterative sub-space methods~\cite{Helgaker2014}. 
The single-reference formulation has the disadvantage to be non-variational and has been shown to fail in the limit of strongly correlated regimes of diatomic molecules such as N$_2$ for which static correlations (with their multi-reference character) are known to play a decisive role~\cite{VanVoorhis2000,Bulik2015,Zhao2016, gomez2016recoupling}.
While a variational version of CC was proposed~\cite{VanVoorhis2000}, the high computational cost associated to the numerical optimization of the CC parameters has limited its applications.

Interestingly, the unitary variant of CC (UCC)~\cite{Hoffmann1988, Kutzelnigg1991, Cooper2010, Evangelista2011} can naturally be mapped to a quantum circuit for the preparation of corresponding wavefunction in a digital quantum computer~\cite{peruzzo_variational_2014, omalley_scalable_2016, Romero2017a, Barkoutsos2018, Moll2018, kuhn2019, bauman2019quantum, evangelista2019exact}. 
Moreover, variational approaches and in particular the Variational Quantum Eigensolver (VQE) algorithm ~\cite{peruzzo_variational_2014} in combination with truncated wavefunction expansions (with polynomial number of parameters) appear, at present, the most promising way of solving chemistry problems on near-term quantum hardware~\cite{Yung2014,Barends2015,mcclean_theory_2016, wang18,Ganzhorn2018}.
Hence, in this work we study the performance of the variational implementation of UCC as a quantum algorithm (q-UCC)~\cite{peruzzo_variational_2014} in computing the ground state of molecules and lattice models for which the classical CC formulation is known to break. 
For the practical implementation of the q-UCC (i.e. reduction of the circuit depth) we also study two variations of the Ansatz, namely pair CC doubles (pCCD)~\cite{Henderson2014, Henderson2015,Bulik2015,Zhao2016} and singlet CC doubles (CCD$0$)~\cite{Bulik2015, gomez2016recoupling}. 
These alternatives were previously developed to address the breakdown of the standard CC theory, in particular when strong correlation effects become important. 
Classically, the CCD and pCCD Ans\"atze were shown to provide more accurate results when combined with orbital optimization (OO) within the Lagrangian CCD-$\Lambda$ formulation~\cite{bozkaya2011quadratically,stein2014seniority,mizukami2019orbital,takeshita2020increasing}.
Here we investigate the implementation of this class of truncated CC expansions into the corresponding quantum algorithms. 
In particular, the OO procedure is embedded as an extension of the VQE algorithm, which we name ooVQE. 
Its performance in terms of accuracy and efficiency is studied in combination with the series of q-UCC Ans\"atze introduced above.
The paper is organized as follows. 
In Section~\ref{sec:theory}, we define the molecular and Hubbard Hamiltonians and recall the theory of the classical CC and UCC methods. 
We also discuss the q-UCC and the quantum equivalents of pCCD and CCD$0$, i.e. q-pUCCD and q-UCCD$0$, as new parametrized wavefunctions.
Section~\ref{sec:methods} describes the implementation of these methods within the framework of the VQE and the ooVQE algorithms.
In Section~\ref{sec:results}, we apply q-UCCSD, q-pUCCD and q-UCCD0 (with and without the use of OO) to the H$_4$, N$_2$, H$_2$O molecules and the one-dimensional Hubbard chain with 6 sites (2 spins), in which strong correlation effects may induce the failure of the standard CC theory.
Conclusions are presented in Section~\ref{sec:conclusion}.

\section{\label{sec:theory}Theory}

\subsection{Molecular Hamiltonian}
Within the Born-Oppenheimer approximation, the non-relativistic molecular Hamiltonian in 
second quantization is given by
\begin{align}
\hat H &=\sum_{rs} \langle r |\hat h| s \rangle \, \sum_{\sigma \tau} \hat{a}^{\dagger}_{r,\sigma} \hat{a}_{s,\tau} \label{eq:H_sec_quant_expl}  \\ &+
\frac{1}{2}\sum_{rstu}\langle r s |\hat g| t u \rangle \,   \sum_{\sigma \tau \nu \mu} \hat{a}^{\dagger}_{r,\sigma} \hat{a}^{\dagger}_{s, \tau} \hat{a}_{u, \nu} \hat{a}_{t, \mu} + E_{NN} , \notag
\end{align}
where
the one-electron integrals, $\langle r |\hat h| s \rangle$, and the two-electron integrals, $\langle r s |\hat g| t u \rangle$, are given in Appendix A; $\hat{a}_{r, \sigma}^{\dagger}$~($\hat{a}_{r, \sigma}$) represent the fermionic creation~(annihilation) operators for electrons in HF spin-orbitals, $\phi_{r,\sigma}(\vec{r})$ with spatial component (molecular orbitals, MOs) $\phi_r(\vec{r})$ and spin $\sigma \in \{\uparrow, \downarrow\}$.
$E_{NN}$ describes the nuclear repulsion energy.
Here and for the remainder of this work we use indices $r,s,t,u$ to label general MOs; $i,j,k,l$ for occupied MOs; $m,n,p,q$ for virtual ones; $\sigma, \tau, \mu, \nu$ for spin components of spin-orbitals. The same spatial orbitals are used for both spin-up and spin-down spin-orbitals.

\subsection{Hubbard Hamiltonian}

The repulsive $N$-site Fermi Hubbard Hamiltonian is defined as
\begin{align}\label{Hubbard}
	\hat{H} =& -t \sum \limits_{r,\sigma}^{N, \{\uparrow, \downarrow\}}(\hat{a}^\dagger_{r+1,\sigma}\hat{a}_{r,\sigma}+\hat{a}^\dagger_{r,\sigma}\hat{a}_{r+1,\sigma}) \\
	&+ U \sum\limits_{r}^{N}\hat{n}_{r,\uparrow}\hat{n}_{r,\downarrow} , \notag
\end{align}
where $t$ is the energy associated with electron hopping,  $U$ is the on-site electronic repulsion and $\hat{n}_{r,\sigma}$ is the number operators $\hat{a}_{r,\sigma}^{\dagger} \hat{a}_{r,\sigma}$. 
The index $r$ labels the lattice sites each of which is divided into two sub-sites for spin up ($\uparrow$) and down ($\downarrow$). 
This implies that the representation of this Hamiltonian on a lattice requires at least $2N$ qubits.

\subsection{Classical Coupled Cluster and Unitary Coupled Cluster Ans\"atze}

A non-linear parametrization of the system wavefunction is given by the CC Ansatz 
\begin{equation}
|\Psi(\vec{\theta}) \rangle =e^{\hat{T}(\vec{\theta})} | \Phi_0 \rangle ,
\label{eq:cc}
\end{equation}
where $| \Phi_0 \rangle$ is the Hartree-Fock state, $\vec{\theta}$ is the CC amplitudes vector and $\hat{T}(\vec{\theta})$ is the full excitation operator,  defined as
\begin{equation}
\hat{T}(\vec{\theta}) = \sum^{n}_{k=1}\hat{T}_k(\vec{\theta}),
\label{eq:T}
\end{equation} 
with $\hat{T}_k(\vec{\theta})$ being the excitation operator of $k^{th}$ order. 
The calculation of the CC amplitudes $\vec{\theta}$ is commonly performed self-consistently, solving the projective CC equations~\cite{Helgaker2014} but may lead to non-variational energies.

The UCC Ansatz defined as
\begin{equation}
|\Psi(\vec{\theta}) \rangle =e^{\hat{T}(\vec{\theta})-\hat{T}^{\dagger}(\vec{\theta})} | \Phi_0 \rangle.
\label{eq:ucc}
\end{equation}
 is of particular importance for quantum computing since the $e^{\hat{T}(\vec{\theta})-\hat{T}^{\dagger}(\vec{\theta})}$ operator is unitary and therefore can be straightforwardly implemented as a quantum circuit.
We stress that the variational UCC method is different to variational CC (vCC) and shows large deviations when strong electron correlation is involved~\cite{harsha2018difference}.

\begin{figure}[ht!]
	\centering
	\includegraphics[width = 1\columnwidth]{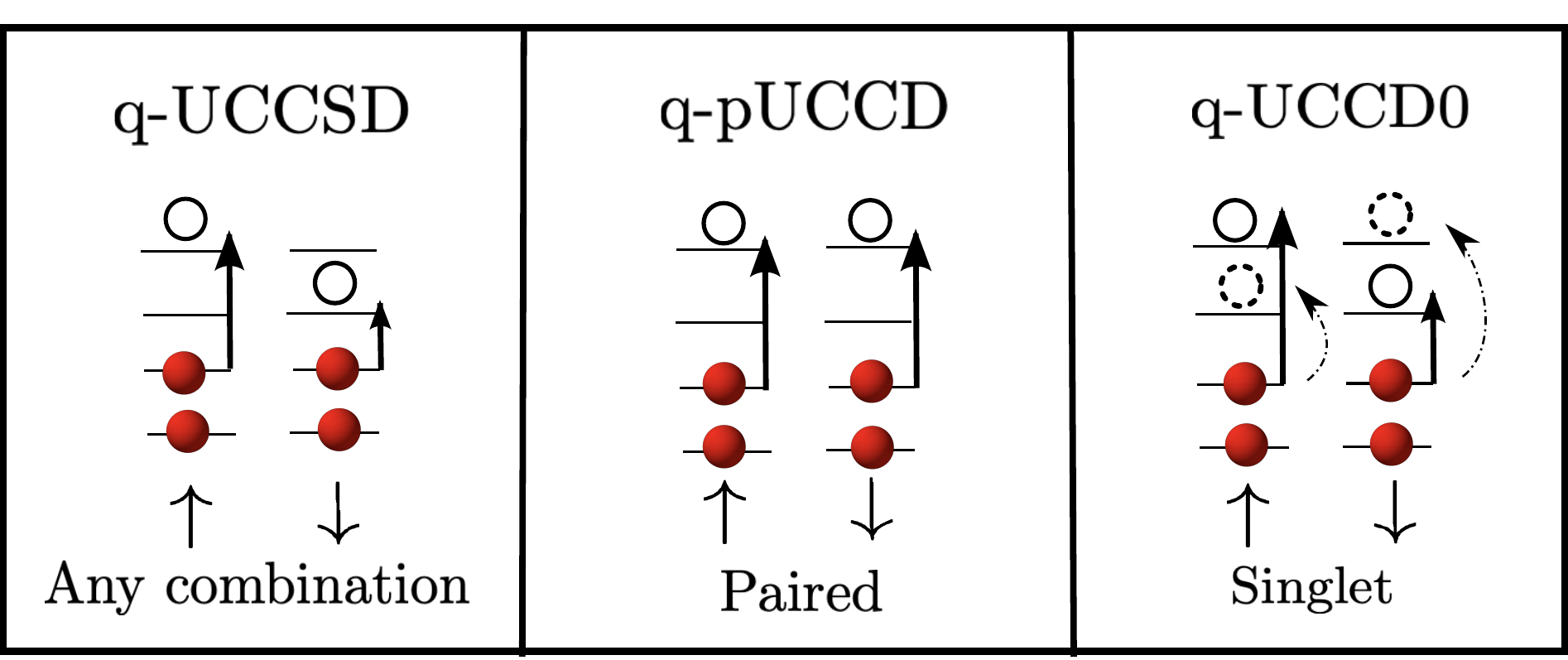}
	\caption{\em Sketch of the possible two-body excitations in different implementations of the q-UCC Ansatz. In the q-UCCSD approach (left-hand panel), double excitations can involve any pair of occupied and virtual orbitals (shown is a double excitation composed by an excitation in the spin-up manifold and one in the spin-down manifold). In q-pUCCD (central panel), double excitations within the spin-up and spin-down MOs are forced to occur in pairs, while in q-UCCD0 (right-hand panel), double excitations indicated with solid and dashed lines are associated to the same amplitudes and therefore only one of the two is explicitly included in the $\hat{T}_2$ operator.}
	\label{fig:advanced_ucc}
\end{figure}

\subsection{Quantum Unitary Coupled Cluster Singles Doubles: q-UCCSD}
\label{subsection:q-UCCSD}

The quantum implementation of the UCC Ansatz, named q-UCC, requires the use of the Trotter formula which allows us to construct the Ansatz with a sequence of quantum gates. 
For two general operators $\hat A$  and $\hat B$, the formula reads
\begin{equation}
e^{(\hat A+ \hat B)} = \lim_{n \rightarrow \infty}\left(
e^{\frac{\hat A}{n}} e^{\frac{\hat B}{n}}
\right)^n + \mathcal{O}(1/n)\, .
\label{eq:trotter}
\end{equation}
For $n=1$ and with restriction to single and double excitations, the q-UCCSD operator $e^{\hat T(\vec{\theta})-\hat T^{\dagger}(\vec{\theta})} $ can be expressed as
\begin{align} 
& e^{\hat{T}_1(\vec{\theta})-\hat{T}_1^{\dagger}(\vec{\theta})}e^{\hat{T}_2(\vec{\theta})-\hat{T}_2^{\dagger}(\vec{\theta})}   \  
\approx \prod_{mi, \sigma}e^{\theta_{m_{\sigma}i_{\sigma}}(\hat{a}_{m, \sigma}^{\dagger} \hat{a}_{i,\sigma}-\hat{a}_{i, \sigma}^{\dagger} \hat{a}_{m, \sigma})} \label{eq:UCCSD_trotter} \notag  \\ 
& \times \prod_{mnij, \sigma\tau}e^{\theta_{m_{\sigma}n_{\tau}i_{\sigma}j_{\tau}}(\hat{a}_{m, \sigma}^{\dagger}\hat{a}_{n,\tau}^{\dagger} \hat{a}_{i,\sigma}\hat{a}_{j,\tau}-\hat{a}_{j, \tau}^{\dagger} \hat{a}_{i, \sigma}^{\dagger}\hat{a}_{n, \tau}\hat{a}_{m, \sigma}) }. 
\end{align}

In ref.~\cite{Barkoutsos2018} we showed that a single Trotter step is sufficient in the VQE approach to reach the ground state energy within chemical accuracy (i.e., with an error less than $1$ kcal/mol or $1.6\cdot10^{-3}$ Hartree) for the H$_2$ molecule.
For the case of q-UCCSD in small systems, there is evidence~\cite{Barkoutsos2018} that the error can be
`absorbed' and distributed over the entire set of q-UCCSD parameters during the VQE optimization.
Although in a recent work~\cite{grimsley2019trotterized}, the authors claim that $n=1$ might not be sufficient and $n=2$ with independent variational parameters for the second Trotter step may be necessary, in this work we are investigating alternative solutions that do not imply an increase in the circuit depth. 

The implementation of Eq.~\eqref{eq:UCCSD_trotter} in a quantum circuit requires the mapping of the fermionic operators to the qubit operators, which is accomplished using the so-called fermion-to-qubit transformations (e.g. the Jordan-Wigner~\cite{Jordan1928}, the parity or the Bravyi-Kitaev~\cite{Bravyi2000} transformations).

\subsection{Variants of Quantum Unitary Coupled Cluster Ansatz}
\label{subsec:subsection_CC_approx}
The q-UCCSD operator (Eq.~\eqref{eq:UCCSD_trotter}) can be approximated by neglecting classes of excitations from the $\hat{T}_2(\vec{\theta})$ operator.
We present two tuned quantum versions of the original CC Ansatz that have shown interesting results in applications to systems in strongly correlated regimes~\cite{Bulik2015}.

\subsubsection{Quantum (Orbital-Optimized) Pair Unitary Coupled Cluster Doubles: q-(oo-)pUCCD}
\label{subsec:subsection_ooCC_appro}

Within the pCCD approach  electrons with opposite spins are only allowed to undergo the same type of excitation (pair) between occupied and virtual HF orbitals. 
More precisely, the double excitation operator $\hat{T}_2(\vec{\theta})$ becomes (see  Fig.~\ref{fig:advanced_ucc})
\begin{equation}
\hat{T}_2^{\rm{pCCD}}(\vec{\theta})=\sum_{mi} \theta_{m_{\uparrow}m_{\downarrow}i_{\uparrow}i_{\downarrow}}\hat{a}^{\dagger}_{m,{\uparrow}}\hat{a}^{\dagger}_{m,{\downarrow}}\hat{a}_{i,{\uparrow}}\hat{a}_{i,{\downarrow}}.
\label{eq:pucc}
\end{equation}
The full pCCD excitation operator then reads 
\begin{equation}
\hat{T}(\vec{\theta})=\hat{T}_2^{\rm{pCC}}(\vec{\theta}) .
\label{eq:t_pccsd}
\end{equation}
Recently, 
Lee et al.~\cite{Lee2018} demonstrated the wavefunction Ansatz related to pCCD, named k-UpCCGSD, can be systematically improved by applying $k$-times the trotterized UpCCGSD operator (which includes generalized single and pair double excitations) with independent amplitudes. 

Classically, the pCCD approach showed an excellent performance when used with OO~\cite{stein2014seniority}. 
In short, the oo-pCCD approach consists of two main steps: (\textit{i}) optimization of orbital rotations using a unitary operation $R=e^{-\kappa}$, where $\kappa$ is an antihermitian matrix with rank equal to the number of MOs~\cite{bozkaya2011quadratically,bozkaya2013orbital}, and (\textit{ii}) the optimization of the pair $\vec{\theta}$ amplitudes associated to the $\hat{T}_2^{\rm{pCC}}$ operator defined in Eq.~\eqref{eq:pucc}.   

Inspired by the classical implementation, we propose a variant of the pCCD approach in which the rotation operator  $R=e^{-\kappa}$ is directly applied to the orbitals instead of acting on the molecular Hamiltonian in second quantization (see Appendix B).
In this way, the matrix $R$ induces simply a change in the Hamiltonian coefficients in Eq.~\eqref{eq:H_sec_quant_expl} according to $\phi'_s=\sum_{r} R_{sr} \phi_r$, where $\{\phi_r\}_{r=1}^N$ represents the initial set of one-electron MOs and $\{\phi'_r\}_{r=1}^N$ the rotated one.
The main advantage of this approach lies in the fact that the gate operations associated to the single excitation operator $\hat{T}_1$ can be replaced by the optimization of the matrix elements of the antihermitian matrix $\kappa$ followed by a re-evaluation of the Hamiltonian matrix elements. 
We name this new method quantum orbital-optimized pair UCCD (q-oo-pUCCD) and give an in-depth study in Appendix C.

\subsubsection{Quantum (Orbital-Optimized) Singlet Unitary Coupled Cluster Doubles (Full): q-(oo-)UCCD0(-full)}\label{sec:singlet}
The q-pUCCD approach does not correlate more than 2 spatial MOs at a time, especially when the restricted HF (RHF) is used to generate the MOs. Another approximation can be used to overcome this limitation while reducing substantially the number of terms in the $\hat{T}_2$ operator.
In the singlet CC method, CCD$0$~\cite{Bulik2015}, the double excitation operator is split into singlet $\hat{T}_2^{0}(\vec{\theta})$ and triplet $\hat{T}_2^{1}(\vec{\theta})$ components
\begin{equation}
\hat{T}_2(\vec{\theta})=\hat{T}_2^{0}(\vec{\theta})+\hat{T}_2^{1}(\vec{\theta}) ,
\label{eq:t2_separated}
\end{equation}
where
\begin{subequations}
	\begin{align}
	\label{eq:t1_singlet}
	\hat{T}_2^{0}(\vec{\theta}) &=  \, \sum_{mnij} \frac{\alpha_{mnij}}{2} \, ( \hat{a}^{\dagger}_{m,\downarrow}\hat{a}^{\dagger}_{n,\uparrow}  + \hat{a}^{\dagger}_{n,\downarrow}\hat{a}^{\dagger}_{m,\uparrow} \, )\\ & \times \, (\hat{a}_{j,\uparrow} \, \hat{a}_{i,\downarrow} + \hat{a}_{i,\uparrow} \, \hat{a}_{j,\downarrow}), \nonumber
	\\ 	\label{eq:t1_triplet}
	\hat{T}_2^{1}(\vec{\theta}) 
	&= \, \sum_{mnij}\frac{\beta_{mnij}}{2}(\hat{a}^\dagger_{m,\uparrow}\hat{a}^\dagger_{n,\uparrow}  \hat{a}_{j,\uparrow}\hat{a}_{i,\uparrow} \, \\ &+ \, \hat{a}^\dagger_{m,\downarrow}\hat{a}^\dagger_{n,\downarrow}  \hat{a}_{j,\downarrow} \, \hat{a}_{i,\downarrow} \nonumber \\ &+ \frac{1}{2} (\hat{a}^\dagger_{m,\downarrow}\hat{a}^\dagger_{n,\uparrow} \,  - \, \hat{a}^\dagger_{n,\downarrow}\hat{a}^\dagger_{m,\uparrow} )(\hat{a}_{j,\uparrow} \, \hat{a}_{i,\downarrow} - \hat{a}_{i,\uparrow} \, \hat{a}_{j,\downarrow})) \nonumber \, .
	\end{align}
\end{subequations}
Using the symmetry under interchange of indices for $\hat{T}_2^{0}$ and the anti-symmetry for $\hat{T}_2^{1}$, the coefficients $\vec \alpha, \vec \beta$ can be related to the original parameters $\vec \theta$ as follows
\begin{align}
\alpha_{mnij}&=	\alpha_{nmij} = 	\alpha_{mnji} =	\alpha_{nmji}= \frac{\theta_{mnij}+\theta_{nmij}}{2}, \label{eq:anlges_sym0}
 \\
\beta_{mnij}&= -\beta_{nmij}= -\beta_{mnji} = \beta_{nmji} = \frac{\theta_{mnij} - \theta_{nmij}}{2},
\label{eq:anlges_sym1}
\end{align}
where we also use the relation $\theta_{mnij} = \theta_{nmji} = \theta _{ m_\uparrow n_\downarrow i_\uparrow j_\downarrow}$~\cite{Piecuch1990}.
As a consequence, only one out of the four cases in~\eqref{eq:anlges_sym0} can be considered for each set of indices  ($m,n,i,j$).
This subset of indices is named $\Omega$.
When using the RHF approach, the spatial MOs involved in such excitations are the same for both spins, adding an extra symmetry to further reduce the number of excitation operators.
We define the q-UCCD$0$ excitation operator (see Fig.~\ref{fig:advanced_ucc}), based on subset $\Omega$, as 
\begin{equation}
\hat{T}(\vec{\theta})=\hat{T}_2^{0,\Omega}(\vec{\theta}),
\label{eq:t_ccsd0omega}
\end{equation}
and 
\begin{align}
\hat{T}_2^{0,\Omega}(\vec{\theta}) &= \, \sum_{mnij \subset \Omega} \theta_{m_{\uparrow}n_{\downarrow}i_{\uparrow}j_{\downarrow}} \, \hat{a}^{\dagger}_{m,\uparrow} \, \hat{a}^{\dagger}_{n,\downarrow}\hat{a}_{j,\uparrow} \, \hat{a}_{i,\downarrow}.
\label{T20_restricttoS}
\end{align}

In addition, we also implement the \textit{full} form of the CCD0 Ansatz proposed by Bulik et al.~\cite{Bulik2015} in which the choice of the indices $(m,n,i,j)$ in the definition of the $\hat{T}_2^{0}(\vec{\theta})$ operator (see Eq.~\eqref{T20_restricttoS}) is not restricted to the $\Omega$ subset 
\begin{equation}
\hat{T}(\vec{\theta})=\hat{T}_2^{0}(\vec{\theta}).
\label{eq:t_ccsd0}
\end{equation}
In the following, we refer to this method as  q-UCCD$0$-full.
Even though the two expansions (q-UCCSD$0$ and q-UCCSD$0$-full) are characterized by the same number of parameters, the number of excitation operators are different in the two cases (in q-UCCD$0$-full Ansatz, up to 4 excitations can be controlled by the same parameter). For this reason, the optimized ground state wavefunction can differ in the two approaches giving rise to different energies and PES shapes.
In Appendix E, we provide the simplest non-trivial example of these Ans\"atze.

The major difference with the triplet operator $\hat{T}_2^{1}(\vec{\theta})$ is that the latter also includes same-spin excitations (i.e., double excitations of the same spin). 
As shown in ref.~\cite{Bulik2015}, when $\hat{T}_2^{1}(\vec{\theta})$ is used instead of $\hat{T}_2^{0}(\vec{\theta})$ for the calculation of the dissociation profile of molecules in strongly correlated regimes, the accuracy for the energy decreases in comparison to CCD0.
Therefore, we will restrict our investigation to the q-UCCD$0$ and q-UCCD$0$-full approaches.

For completeness, we have also combined q-UCCD0 and q-UCCD0-full Ans\"{a}tze with the OO described in Section~\ref{subsec:subsection_ooCC_appro}. 
The corresponding methods are named q-oo-UCCD0 and q-oo-UCCD0-full.

\section{\label{sec:methods}Methods}

\subsection{VQE Algorithm}
\label{subsec:vqe}

The implementation of the q-UCC wavefunction Ansatz in near-term quantum hardware requires the application of techniques for the reduction of the circuit depth in such a way that the overall execution time of the algorithm does not exceed the coherence time of the quantum computer.
To this end, in addition to the methods introduced in Section~\ref{subsec:subsection_CC_approx} we will make use of precision-preserving qubit-reduction schemes proposed in refs.~\cite{Bravyi2017a, moll_optimizing_2016, Setia2017} via their implementation in the Qiskit software platform~\cite{qiskit2019}. 
In all our applications, we use the VQE algorithm~\cite{peruzzo_variational_2014} for the calculation of the ground state energies according to the following steps:
\begin{itemize}
	\item[(\textit{1})] 
	After setting the coordinates, charge and spin multiplicity of the molecule, we perform
	a RHF calculation in the minimal, STO-3G basis set, using the PySCF package~\cite{Sun2018}. 
	\item[(\textit{2})] 
The matrix elements:\ $\langle r | \hat h | s \rangle$ and  $\langle rs | \hat g | tu \rangle$ are then extracted and used to construct the molecular Hamiltonian (Eq.~\eqref{eq:H_sec_quant_expl}) using the parity fermion-to-qubit mapping~\cite{Bravyi2000}.\
	Exploiting the symmetries
	\begin{equation} \label{eq:z2symm}
	    [\hat{H},\hat{N}_{\uparrow}]=[\hat{H},\hat{N}_{\downarrow}]=0 \, ,
	\end{equation}
	we can combine parity mapping with a two-qubit reduction (one of each $\mathbb{Z}_2$ symmetry of the Hamiltonian) without modifying the lower part the energy spectrum (including the ground state), as described in ref.~\cite{Bravyi2017a}. 
	(In Eq.~\eqref{eq:z2symm} $\hat{N}_\sigma$ is the number operator for electrons of spin $\sigma$.)  
	Finally, the frozen-core approximation~\cite{VonBarth1980} is employed to reduce the number of possible single and double excitations and the qubit count.
	\item[(\textit{3})] 
	The qubits can be further tapered off~\cite{Bravyi2017a} by finding the underlying symmetries of the Hamiltonian and using graph-based qubit encodings. The latter applies to the Hamiltonian, the q-UCC operator and the state vector. Further details of the tapering procedure can be found in Appendix F.
	\item[(\textit{4})]
	The trial wavefunction $|\Psi(\vec{\theta}) \rangle$ is generated starting from the HF state $|\Phi_0\rangle$ 
	by applying 
	the cluster operator Eq.\eqref{eq:UCCSD_trotter} chosen among the q-UCC Ans\"atze.
	\item[(\textit{5})] 
    The system energy $\langle \Psi(\vec{\theta}) |\hat{H}|\Psi(\vec{\theta})  \rangle$ is evaluated using the state-vector simulator provided by Qiskit~\cite{qiskit2019}, which uses a matrix representation of the operators in the Hilbert space. 
    \item[(\textit{6})]
Steps (\textit{4}) and (\textit{5}) are repeated until convergence using a classical optimizer.\
    In all our simulations, we employ the Sequential Least Squares Quadratic Programming (SLSQP)~\cite{Kraft1988} optimization algorithm, which in our implementation performs in terms of number of steps equally well as the L-BFGS-B optimizer (shown~\cite{Romero2017a} to be among the best optimizers for VQE used with the q-UCCSD Ansatz). \ 
    In the first iteration, all amplitudes $\vec\theta$ are initialized to a fixed value set to $0.1$. 
    While in principle possible, in this work we do not start the optimization using better guesses such as MP2 amplitudes.
    The convergence criterion for the energy is set to $10^{-7}$.
\end{itemize}
Note that the calculation in step (\textit{5}) can also be performed through the sampling of the expectation value of the Hamiltonian by repeated executions of the quantum circuit that encodes the trial wavefunction. 
This approach will lead to the same ground state energy as the state-vector simulation at a much larger computational cost. 
All calculations are performed using the VQE algorithm and the SLSQP optimizer as implemented in Qiskit~\cite{qiskit2019}.

\subsection{ooVQE Algorithm}\label{subsec:oovqe}

For the implementation of the OO approaches discussed in Section~\ref{sec:singlet}, we modify the general framework of the VQE algorithm. 
The one- and two-body integrals used to generate the Hamiltonian matrix are modified according to

\begin{align}
  \langle r |\tilde{\hat{h}}| s \rangle &= \sum_{a b}C^{*}_{a r} C_{b s} \langle a |\hat{h}| b \rangle,  \label{eq:h_rotated_oovqe} \\
  \langle p q |\tilde{\hat{g}}| r s \rangle &= \sum_{a b c d}C^{*}_{a p} C^{ *}_{b q} C_{c r} C_{d s} \langle a b  |\hat{g}| c d \rangle, \label{eq:g_rotated_oovqe}
\end{align}
with $C={C}_{\text{RHF}} e^{-\kappa}$  and $\kappa$ is an antihermitian matrix.
The atomic orbital (AO) to MO coefficients matrix is given by $C_{\text{RHF}}$ (where the indices $a, b, c$ and $d$ label the AOs).
In particular, we modify the following two steps in the conventional VQE to obtain the ooVQE algorithm:
\begin{itemize}
	\item[(\textit{2})$^{*}$] 
    Extracted RHF integrals $\langle r | \hat h | s \rangle$ and  $\langle rs | \hat g | tu \rangle$ undergo orbital rotation using Eqs.~\eqref{eq:h_rotated_oovqe} and~\eqref{eq:g_rotated_oovqe}.
	\item[(\textit{6})$^{*}$] 
	In addition to amplitudes $\vec{\theta}$, matrix elements $\vec{\kappa}$ are introduced into the optimization. Their initial value is arbitrarily fixed at $10^{-3}$. At every update of $\vec{\kappa}$ vector by the optimizer the Hamiltonian matrix elements are reconstructed using Eqs.~\eqref{eq:h_rotated_oovqe} and~\eqref{eq:g_rotated_oovqe}.
\end{itemize}

From this point, the ooVQE algorithm proceeds unchanged until convergence as for the conventional VQE approach.
Note that by construction, the Hamiltonians before and after the rotation of the orbitals share the same energy spectrum due to unitarity of applied orbital rotation. 
However, through the optimization of the orbitals, we aim at minimizing the distance between the exact wavefunction and the support specified by any q-UCC Ansatz.
A more detailed description of the algorithm is given in Appendix B.

\subsection{Classical Electronic Structure Calculations}\label{subsec:classical_methods}

The CCSD calculations are performed using the PySCF package~\cite{Sun2018}.
For comparison, we also report results obtained using the pair orbital-optimized M{\o}ller-Plesset method (pOMP2) and its reference OMP2 as implemented in Psi4~\cite{turney2012psi4, smith2018psi4numpy}. 
The implementation of the pOMP2 approach follows closely the prescriptions described in ref.~\cite{bozkaya2011quadratically, stein2014seniority}, where all the amplitudes with non-zero seniority~\cite{chen2015seniority} (the number of singly occupied orbitals in a determinant or an orbital configuration), which are not associated to the paired double excitations are eliminated from MP2 Lagrangian equations.

\section{\label{sec:results}Results and Discussion}

\subsection{How Different is q-UCCSD from its Classical UCCSD Equivalent?}

In the Section~\ref{subsection:q-UCCSD} as well as in ref.~\cite{Barkoutsos2018}, we pointed out that due to the use of the Trotter approximation (Eq.~\eqref{eq:UCCSD_trotter}) with $n=1$ the q-UCCSD approach cannot be considered a one-to-one map of the original corresponding classical algorithm, UCCSD. 
Despite this approximation, the q-UCCSD method can reproduce the correct ground state energy when optimized with the VQE algorithm.

In this Section, we investigate the evolution of UCCSD and q-UCCSD variational parameters $\vec \theta$ and the corresponding wavefunction, $|\Psi(\vec \theta)\rangle$, for the case of the \ce{H$_4$} molecule, a system that will be further studied in Sec.~\ref{sec:h4} to assess the quality of the q-UCCSD approach and its approximations (q-(oo)-pUCCD, q-(oo)-UCCD0 and q-(oo)-UCCD0-full).

\begin{figure}[ht]
\centering
\includegraphics[width=1.\linewidth]{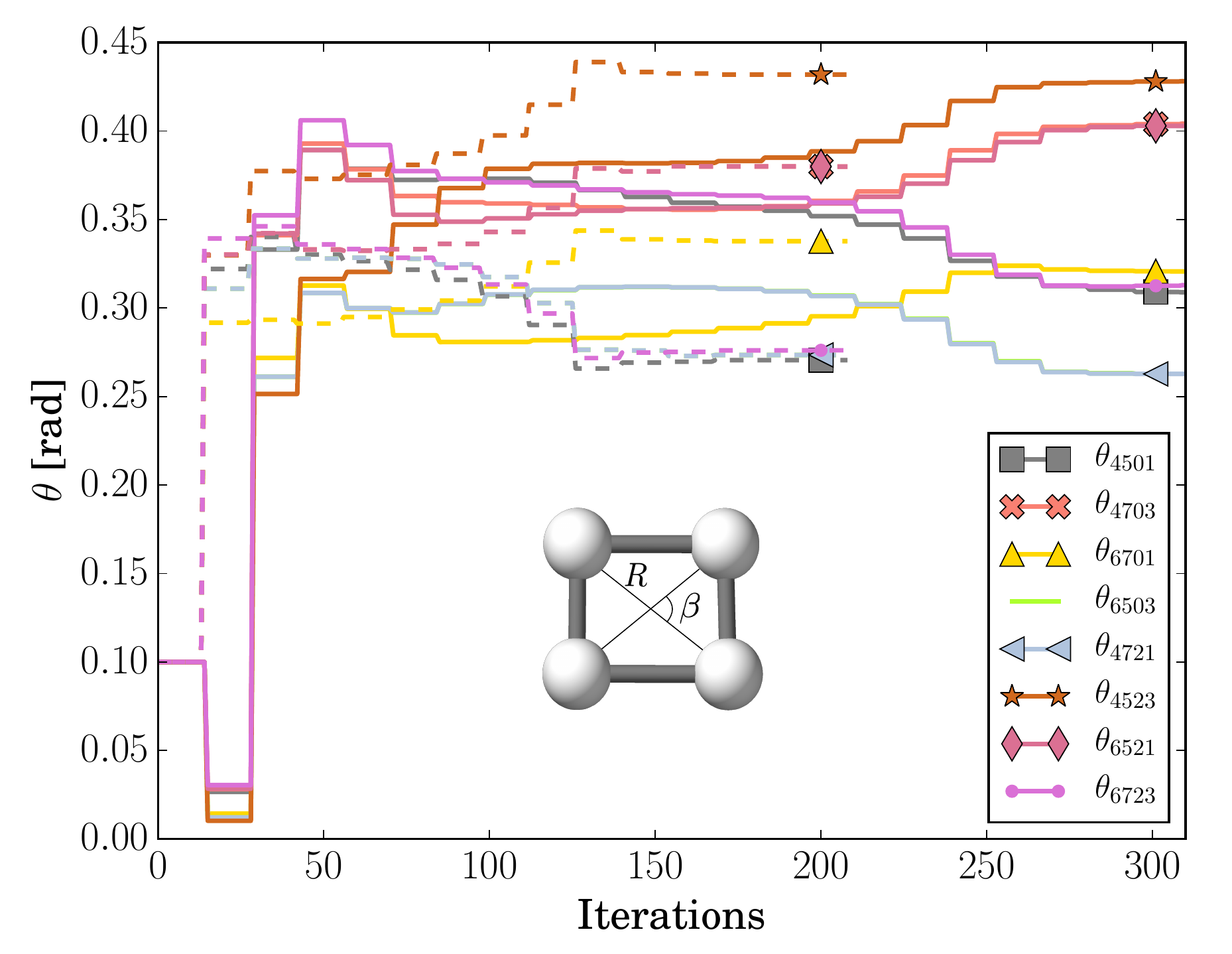}
\caption{\it Evolution of the most relevant VQE parameters for the $H_4$ molecule during optimization (STO-3G basis, 8 qubits, 4 electrons, R= 1.735 \AA, $\beta=85^{\circ}$) using the SLSQP optimizer (tol $=10^{-7}$).
	The q-UCCSD results (solid line) are compared with the ones obtained using the classical UCCSD approach (dashed line) which is equivalent to the exact Trotter decomposition. The angle $\theta_{m_{\sigma}n_{\tau}i_{\sigma}j_{\tau}}$ corresponds to the excitation $\hat{a}^{\dagger}_{m,{\sigma}}\hat{a}^{\dagger}_{n,{\tau}}\hat{a}_{j,{\sigma}}\hat{a}_{i,{\tau}}-\hat{a}^{\dagger}_{i,{\tau}}\hat{a}^{\dagger}_{j,{\sigma}}\hat{a}_{n,{\tau}}\hat{a}_{m,{\sigma}}$.
	}
	\label{fig:H4_angles}
\end{figure}

\begin{figure}[ht]
\centering
\includegraphics[width=1.\linewidth]{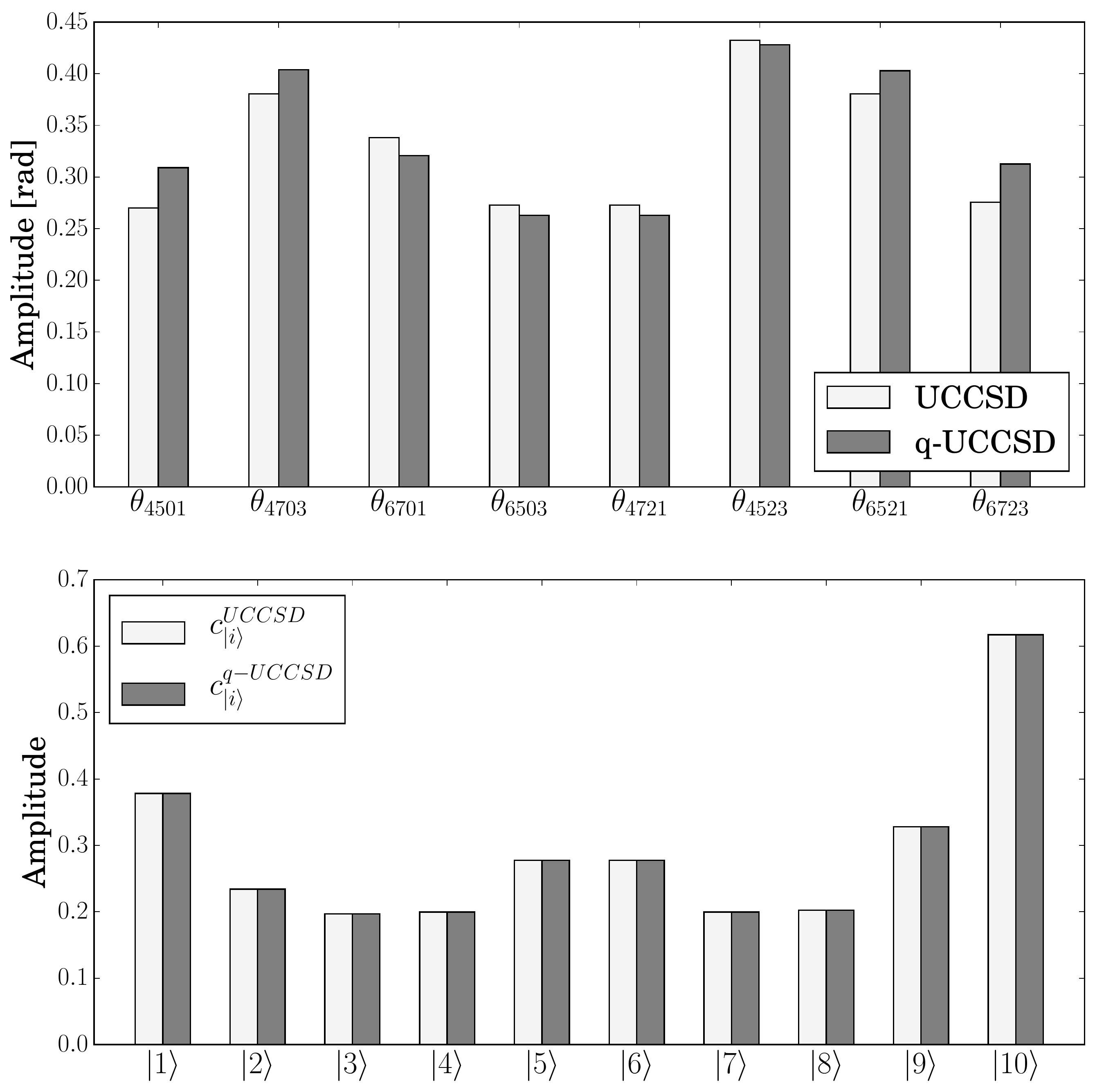}
	\caption{\it \textbf{Upper panel: } Converged $\theta_{mnij}$ parameters corresponding to the optimization profiles in Fig~\ref{fig:H4_angles}. The differences are due to the Trotter approximation used for q-UCCSD. \textbf{Lower panel: } Corresponding amplitudes in Eq.\eqref{eq:fci} for the first 10 dominant configurations (out of 256) of the ground state wavefunction.
	States $| 1 \rangle, ...,| 10 \rangle $ denote the electron configurations 
	$|00001111 \rangle$, 
	$|00110011 \rangle$, 
	$|00111100 \rangle$, 
	$|01100110 \rangle$, 
	$|01101001 \rangle$, 
	$|10010110 \rangle$,
	$|10011001 \rangle$, 
	$|11000011 \rangle$ 
	$|11001100 \rangle$,
	$|11110000 \rangle$, respectively (spin-orbitals are ordered from lowest to highest energies (left to right) with up (down) spins in the even (odd) positions).}
	\label{fig:H4_angles_histo}
\end{figure}

Fig.~\ref{fig:H4_angles} shows the evolution of the most relevant variational parameters along the optimization path for both the exact implementation of the q-UCCSD (with no Trotter approximation, obtained by matrix exponentiation of the exact cluster operator $(\hat{T}-\hat{T}^{\dagger})$) and the approximated algorithm (Trotter expansion with $n=1$, see Eq.~\eqref{eq:UCCSD_trotter}). 
Despite the small differences in the paths followed by the two approaches, the converged $\vec{\theta}$ values (Fig.~\ref{fig:H4_angles_histo}, upper panel) produce the same ground state wavefunction, characterized by the same expansion coefficients $c^{\sigma}_{mi}$ and  $c^{\sigma \tau}_{mnij}$ defined in linearly parametrized wavefunction
\begin{align}
|\Psi(\vec{c}) \rangle =(1&+\sum_{mi,\sigma}c^{\sigma}_{mi}\hat{a}^{\dagger}_{m, \sigma}\hat{a}_{i, \sigma} \notag \\
&+\sum_{mnij, \sigma \tau}c^{\sigma \tau}_{mnij}\hat{a}^{\dagger}_{m, \sigma}\hat{a}^{\dagger}_{n, \tau}\hat{a}_{i, \sigma}\hat{a}_{j, \tau})
| \Phi_0 \rangle.
\label{eq:fci}
\end{align}
The weights of the first ten dominating configurations in Eq.~\eqref{eq:fci} contributing to the ground state wavefunction are shown in Fig.~\ref{fig:H4_angles_histo} (lower panel) for the `exact' UCCSD and q-UCCSD approaches. The agreement is good with a maximal deviation of the order  of $10^{-7}$.
This analysis confirms the accuracy of the q-UCCSD approach based on  Eq.~\eqref{eq:UCCSD_trotter} within the VQE framework.

\subsection{q-UCC for Systems in the Strongly Correlated Regime}

In the following, we will analyze the performance of the q-UCC Ans\"atze described in Section~\ref{subsec:subsection_CC_approx} within VQE and ooVQE algorithms for a series of molecular systems and the periodic  Fermi-Hubbard model.
A summary of all results is given in Tab.~\ref{tab:results}.

In order to facilitate the comparison between the different methods, we shifted all energy profiles to match the same reference value at the equilibrium (i.e., the minimum energy) geometry. The values of the shifts are also reported in Tab.~\ref{tab:results} and in the legends of the figures~\ref{fig:H4_all}, \ref{fig:H2O_all} and \ref{fig:N2_all}.

As a measure of the quality of the dissociation/distortion profiles, we use the non-parallelity error (NPE)~\cite{dutta2003full,abrams2004full} defined as the difference between the maximum and minimum error, over the entire energy profile, with respect to exact diagonalisation of the Hamiltonian in the given basis set.

\begin{table*}[ht]
\centering
\caption[]
{
\textit{Results of the different q-UCC Ans\"atze within VQE and ooVQE algorithms. 
Description of the abbreviations: 
Hub.: Hubbard;
No.: number; 
SQG: single qubit gates, u1,u2,u3 corresponding to one-,two-, and three-parameters one-qubit gates (details are presented in ref.~\cite{qiskit2019});
TQG: two-qubit gates, CNOT gates;
par.: parameters, include single and double excitations with "+" indicating the number additional parameters of $\kappa$ matrix used for OO;
en. evals: number of energy evaluations in VQE; 
VQE it.: number of VQE iterations;
En. sh.: energy shift;
En. err.: average energy error on the ground state;
NPE: non-parallelity error;
Err. bar.: error in the dissociation barrier;
Imp. OO: improvement on the absolute energy due to OO;
Errors in the dissociation barrier (difference between maximum and minimum energy) and NPEs are calculated with respect to the exact diagonalisation.
When tapering is applied, the number of gates is given in Table~\ref{tab:results_tapering} of Appendix D.
The uncertainties were calculated with the standard error $\sigma / \sqrt{N}$ where $N$ is the number of data points.
}}
\label{tab:results}
\tiny
    \begin{tabularx}{\textwidth}{m{1.3cm}m{1.95cm}m{1.95cm}m{1.95cm}m{1.95cm}m{1.95cm}m{1.95cm}m{1.95cm}m{1.95cm}ccccccccc}
\hline \hline
 &    & q-UCCSD & q-UCCD0 & q-pUCCD & q-UCCD0-full  & q-oo-UCCD0 & q-oo-pUCCD & q-oo-UCCD0-full\\
\hline
    & No.\ SQG                 &       	128/464/377	&		 	 	 	72/320/576	&	32/144/114	 	&	 	128/576/450  &  72/320/576       & 	32/144/114        & 	128/576/450   
      \\
    & No.\ TQG           &    	792	&	 	576	&	256	 	&	 1024 &   576	        & 256        & 1024     \\ 
    & No.\ par.               &       	14	&		 	 	 	9	&	8	&	 	9   &   9+6        & 8+6        & 9+6   \\
    & No.\ en. evals            &       	599	&		 	 	 	196	&	181	 	&	 	224 &   376        & 225        & 285   	 \\
H$_4$      & No.\ VQE it.              &       	23	&   	9	&	8	& 	 9    &   9       & 8       & 9    \\
    &  En. sh. [mHa]            &      -0.2	&		 	 	 	-115.8	&	-264.3	 	&	 	-0.7	 &   -1.6        & -2.5        &   -0.6    \\
(6 qubits)    & En. err. [mHa]             &       	0.5 $\pm 0.2$	&		 	 	 	115.5 $\pm 0.1$	&	266.6 $\pm 1.3$	 	&	 	1.3 $\pm 0.4$	 &   1.2 $\pm 0.1$        & 3.6 $\pm 0.4$        & 1.2 $\pm 0.4$   \\
    & NPE [mHa]            &       	0.9	&		 	 	 	0.5	&	4.3	 	&	 	1.4 &   1.5        & 2.2        &   1.4    \\
    & Err.\ bar. [\%]             &       58	&		 	 	 	34	&	265 &    90      & 89        & 140   &	89 \\
    & Imp. OO [\%]            &       -		&	-		& 	 -         &   -      & 99   & 98 &      6     &	    \\
\hline
    & No.\ SQG                    &       	840/8360/3126	&		 	 	 	240/1968/916	&	64/528/244 	&	 512/4224/1924 &   	240/1968/916	&	64/528/244 	&	 512/4224/1924       \\
    & No.\ TQG           &    	11932	&		 	 	 	3000	&	800	 	&	 	                                      6400   &   3000	&	800	 	&   6400     \\ 
    & No.\ par.                 &       	58	&		 	 	 	30	&	8	 	&	 	30  &   30+15        & 8+15        & 30+15    \\
    & No.\ en. evals                &       	2053	&		 	 	 	453	&	73	 	&	 	529  &   1935        & 743        & 912   \\
H$_2$O       & No.\ VQE it.               &       	25	&		 	 	 	14	    &	    7	 	&	 	16  &   41        & 29        & 19    \\
    &  En. sh. [mHa]           & 	 	-0.2	&	-66.9	 	&	 	-117.0	&	-6.29  &   -8.25        &  -11.3        &   -4.0    \\
(10 qubits)    & En. err. [mHa]              &    2.2 $\pm 0.7$	&		 	 	 	68.5 $\pm 0.4$	&	149.4 $\pm 9.2$	 	&	 11.5	$\pm 1.8$  &   10.8 $\pm 1.2$       & 18.9 $\pm 2.8$          & 10.9 $\pm 2.5$ \\
    & NPE [mHa]           &       	5.2     &		 	 	 	2.9	&	57.6	 	&	 11.0	 &   7.1        & 15.9        & 13.9    \\
    & Err.\ bar. [\%]              &       	7.9	&		 	 	 	4.2	& 96  &    18     &     9        & 26        & 23    \\
    & Imp. OO [\%]     &    -  	&	-	&   - 	&   -	 &   87        & 84        & 5    \\
\hline
     & No.\ SQG     &       	1860/22200/7087	&	720/8704/2823	&	120/1456/471	 	&	 	1800/21849/6967   &     720/6272/3910	&	120/1456/471	 	&	 	1800/21849/6967    \\
     & No.\ TQG              &    	30928	&		 	 	 	12192	&	2032	    &     11792	&	11792	 	&	              2032  & 11792  \\ 
     & No.\ par.                  &       	225	&		 	 	 	90	&	15	 	&	 	90  &     90+28        & 15+28        & 90+28 \\
     & No.\ en. evals                &       	6025	&		 	 	 	2140	&	179	 	&	 	2888  &     4241        & 915        & 2535  \\
N$_2$      &  No.\ VQE it.             &       	25	&		 	 	 	57	&	58	 	&	 	17  &     30        & 19        & 15  \\
    & En. sh. [mHa]           &     -2.9	&		 	 	 	-71.6	&	-78.9	 	&	 -22.1     &     -66.7	       & -73.7        & -18.2  \\
(14 qubits)     & En. err. [mHa]              &       	5.4$\pm 0.3$	&		 	 	 	63.4$\pm 1.2$ &	84.1$\pm 1.5$	 	&	 	30.1$\pm 2.0$  &     58.5 $\pm 1.2$        & 78.4 $\pm 1.5$        & 25.1$\pm 2.0$  \\
    & NPE [mHa]              &       	16.0	&		 	 	 	60.4	&	64.6 	&	 	91.6	&     60.4	        & 64.6        & 91.6  \\
    & Err.\ bar. [\%]             &       	2.75  &		 	 	 	0.58	&	21 	 	&	 		33 &     0.58        & 	21 	 	&	 		33  \\
    & Imp. OO [\%]             &        -        &    -   &		- 	&	- 	&     2        & 3        & 2  	 \\
\hline
 &    & q-UCCSD & q-UCCSD0 & q-pUCCSD &   &  &  & \\
\hline
    &  No.\ SQG                    &       	828/5856/2909	&		 	 	 	468/2880/1565	&	252/1120/733	 &             &         &     &         \\
    & No.\ TQG             &    	9200	&		 	 	 	4640	&	1904 &             &         &  	 &         \\ 
    & No.\ par.                   &       	117	&		 	 	 	72	&	45	 &             &         &    &       	  \\
    & No.\ en. evals                &       	16708	&		 	 	 	7136	&	2461	 	&             &         &    &       	  \\
Hub.  & No.\ VQE it.               &       	140	&		 	 	 	96	&  52	&             &         &    	 	  &           \\
    & En. err.[mHa]          &    	353.4$\pm 78.1$	&		 	 	 	387.9$\pm 82.9$	&	594.5$\pm 117.9$	 &             &         &  	  &       	 \\
 (12 qubits)    & NPE [mHa]             &       	424.1	&		 	 	 	465.4	&	713.4	 	 &         &         &         &       	 	  \\
    & Err.\ bar. [\%]            &       	4.5	&		5.1	& 16.1	&             &         &   &       	 \\
\hline \hline
\end{tabularx}
\end{table*}

\subsubsection{H$_4$ Molecule\label{sec:h4}}
\begin{figure}[ht]
\centering
\includegraphics[width = \columnwidth]{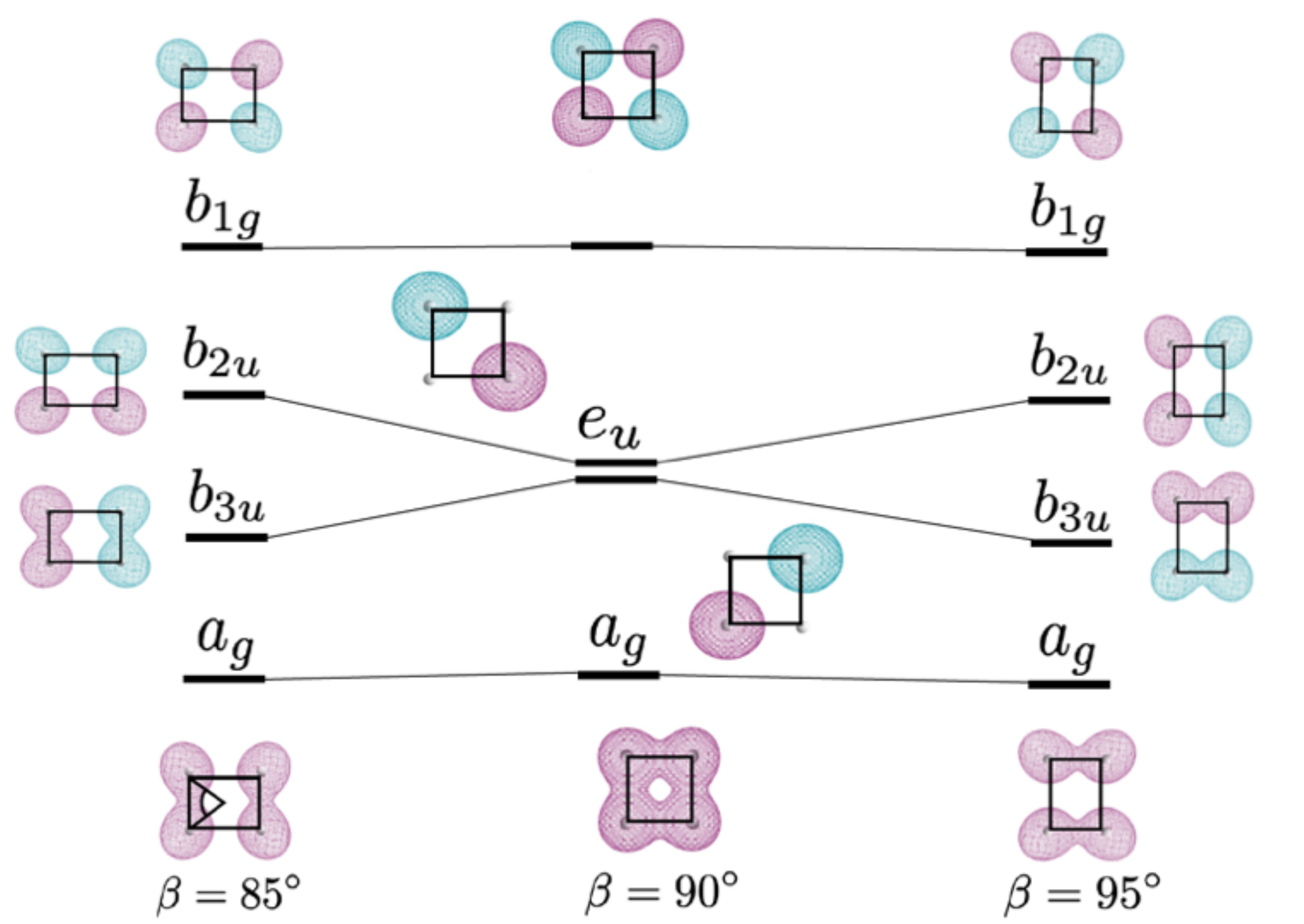}
	\caption{\it Energy diagram for the 4 MOs of the H$_4$ molecule in STO-3G basis set along the deformation profile parametrized by the angle $\beta$ for $R=1.735$~\AA \ and $\beta \in [85^{\circ},95^{\circ}]$.
	}
	\label{fig:advanced_ucc_H4_MO}
\end{figure}

The simplest molecule that exhibits a breakdown of the standard CCSD method is H$_4$ in its planar ring geometry (Fig.~\ref{fig:advanced_ucc_H4_MO}).
While unstable, this molecule was nevertheless extensively used as a benchmark for different computational chemistry methods~\cite{VanVoorhis2000, Genovese2019}.

\begin{figure*}[ht]
\centering
\includegraphics[width=1.\linewidth]{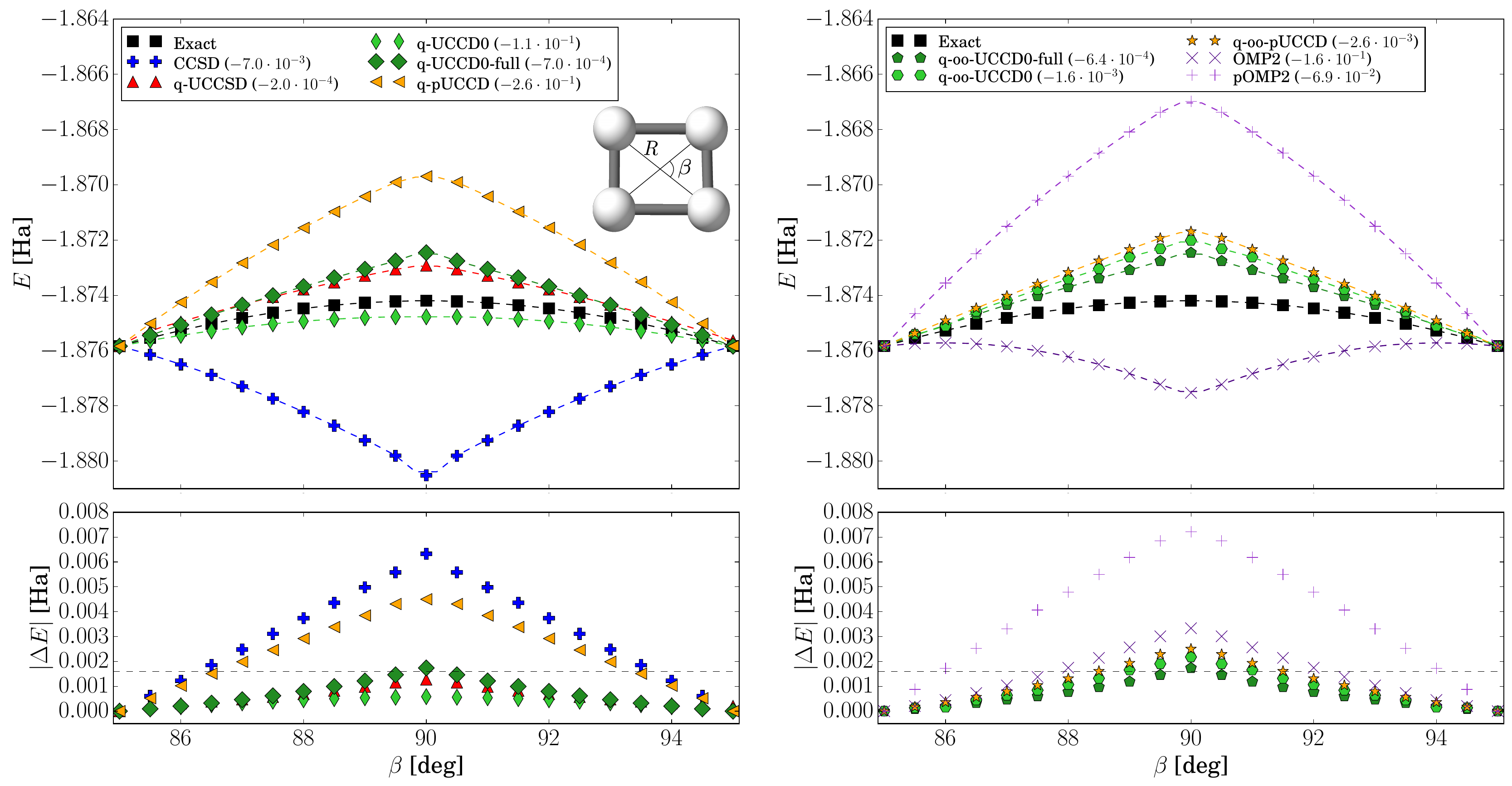}
\caption{
\it \textbf{Upper panels:} Energy profiles of the H$_4$ molecule as a function of the internal angle $\beta$ at $R=1.738$~\AA \space computed using different variants of the classical and quantum CC approaches. 
	All profiles are shifted to match the exact curve at $\beta=85^\circ$. 
The shifts in Hartree are reported within parenthesis next to the acronyms labelling the different approaches (see also Table~\ref{tab:results}).  
	\textbf{Lower panels:}
    Absolute energy differences with respect to the exact profile (obtained with exact diagonalization of the Hamiltonian). 
    The black dashed line corresponds to the chemical accuracy threshold at 1.6 mHa.}
\label{fig:H4_all}
\end{figure*}

We analyze the energy profiles obtained by moving the four hydrogen atoms in a concerted manner along the circumference with radius $R=1.738$~\AA~and with its center coinciding with the center of H$_4$.
All geometries are obtained by varying the angle $\beta$ (Fig.~\ref{fig:H4_all}) from $85^{\circ}$ to $95^{\circ}$, with $\beta=90^{\circ}$ corresponding to the square geometry.

Along this path, the C$_{4v}$ symmetry of the $\beta=90^{\circ}$ geometry is reduced to a C$_{2v}$ ($\beta \neq 90^{\circ}$) symmetry for all other points.
This feature is particularly interesting as the configurations $b_{2u}$ and $b_{3u}$ become degenerate (see Fig.~\ref{fig:advanced_ucc_H4_MO}) at the square  geometry~\cite{Kowalski1998, Kowalski2001}, which is strongly correlated and described by two degenerate molecular states. 

Due to its small size (8 MOs using the STO-3G basis set) and its interesting electronic structure properties, this system is ideal to test the robustness of the different q-UCC-based approaches in the neighbourhood of a multi-configuration point ($\beta= 90^{\circ}$).

Fig.~\ref{fig:advanced_ucc_H4_MO} depicts the changes of the MOs along the deformation path described by the $\beta$ angle, showing the origin of the degeneracy at the square geometry.

In Fig.~\ref{fig:H4_all} (upper panel), we present the energy profiles computed using the classical CCSD algorithm together with the ones obtained using the quantum q-UCCSD, q-UCCD0, q-UCCD0-full and q-pUCCD wavefunction Ans\"atze (left-hand panel) and the corresponding OO forms using ooVQE algorithm (right-hand panel).
Due to the multi-reference character around the $90^{\circ}$-geometry, the classical CCSD approach fails to predict the convexity of the exact energy profile, which is qualitatively wrong with an energy minimum at $90^{\circ}$ instead of a maximum.
All profiles are shifted to match the exact curve at $\beta=85^\circ$. 

Concerning the approximated Ans\"atze, as expected we observe a larger up-shift of the curves in the order of $100$ mHa.  
However, the energy profiles are qualitatively in good agreement with the exact curve in all cases, with the q-UCCD0 Ansatz reproducing the correct profile within an error $<2$ mHa (after the shift). 
The values of the applied shifts for all curves are given in the legend of Fig.~\ref{fig:H4_all} and are summarized in Table~\ref{tab:results}, while the relative energy errors with respect to the exact solution are reported in the lower panel.
The results obtained with the OO procedure described in Section~\ref{subsec:subsection_ooCC_appro} and Appendix B together with the ooVQE algorithm are reported in the right-hand panel of the same figure. 
We observe that for all OO approaches the absolute energy error reduces to values $<4$ mHa
compared to the exact curve. 
In addition, all distortion profiles show the correct qualitative behaviour with a maximum at the square geometry, in contrast to the classical CCSD solution. 
The best results are obtained for the q-oo-UCCD0-full, which gives a profile with a maximal absolute deviation of $1.7$ mHa compared to the reference.
However, note that the computationally efficient q-oo-pUCCD Ansatz, which requires substantially less number of two-qubit gates than all  other methods (see Tab.~\ref{tab:results}) can reproduce qualitatively the correct energy profile.
The q-oo-UCCD0 Ansatz provides very similar results to q-oo-pUCCD except for the error at the top of the barrier at $90^\circ$, which reduces from $140\%$ to $89\%$.
We observe that the OMP2 fails similarly to CCSD to describe the dissociation profile but provides an improvement by reducing the cusp at the square geometry.
The restriction to pOMP2 partially mitigates the errors of OMP2.
However, the size of the barrier between $85^{\circ}$ and $95^{\circ}$ is drastically overestimated.

\subsubsection{H$_2$O Molecule}

Similarly to H$_4$, in the double dissociation of the water molecule by the simultaneous stretching of both OH bonds there are two equally weighted configurations that contribute to the ground state wavefunction~\cite{VanVoorhis2000}.
However, contrary to H$_4$, the non-bonding electron configuration plays a more active role, making the calculation of the ground state energy more challenging for the CC methods.

\begin{figure*}[ht]
\centering
\includegraphics[width=1.\linewidth]{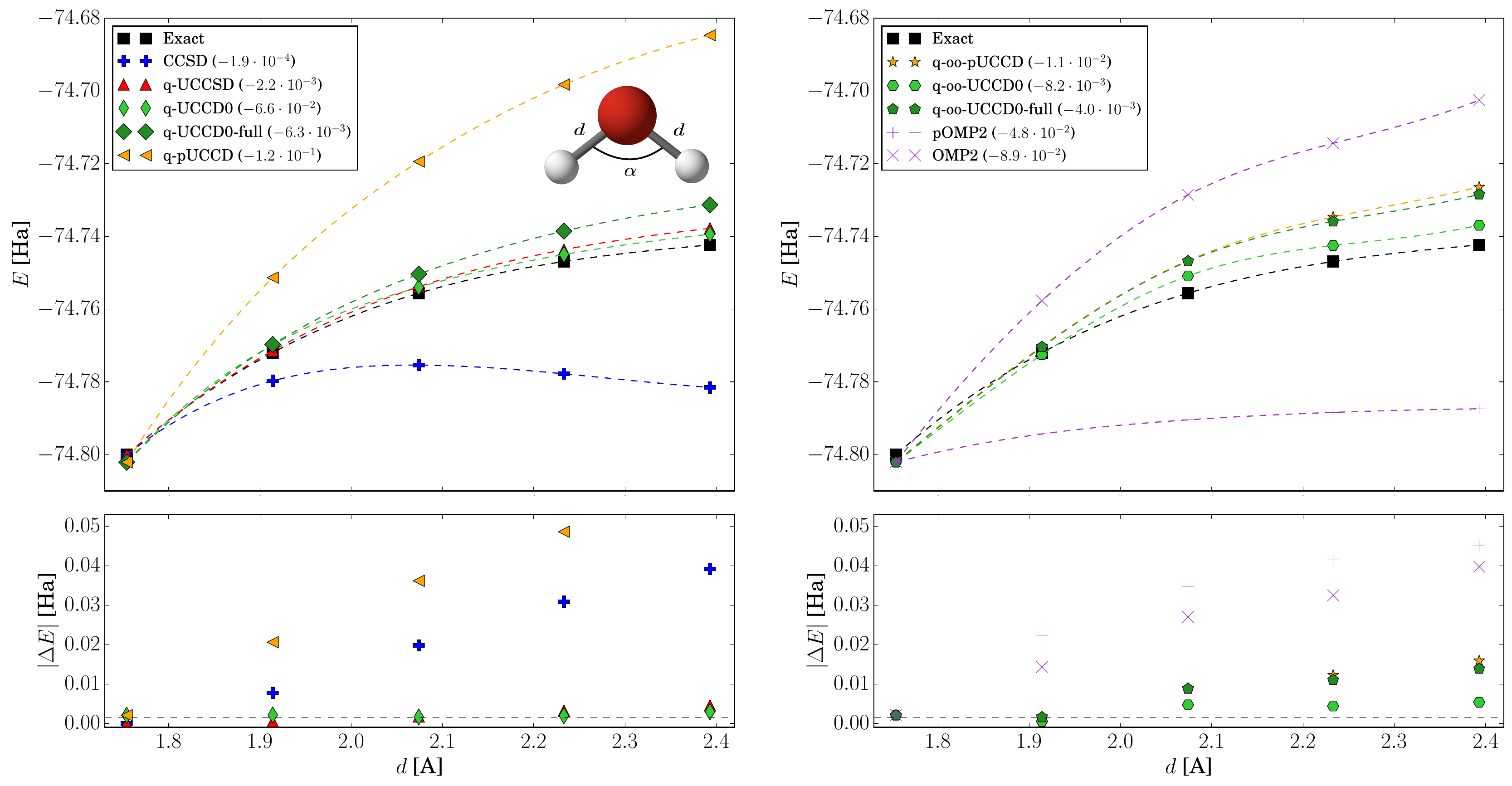}
\caption{
\it \textbf{Upper panels:} 
Double dissociation profiles for H$_2$O computed with different q-UCC Ans\"atze (see legend) at fixed $\alpha= 104.51^{\circ}$.
All profiles are shifted to match the exact curve at a OH distance of $d=1.754$~\AA (see text). 
The shifts in Hartree are reported within parenthesis next to the acronyms labelling the different approaches (see also Table~\ref{tab:results}). 
	\textbf{Lower panels:}  
	Absolute energy differences with respect to the exact profile (obtained upon diagonalization of the Hamiltonian). The black dashed line corresponds to the chemical accuracy threshold at 1.6 mHa. }
\label{fig:H2O_all}
\end{figure*}

The energy profiles for the simultaneous stretching of the OH bonds in the water molecule are shown on the left-hand side of Fig.~\ref{fig:H2O_all} for the CCSD, q-UCCSD, q-UCCD0, q-UCCD0-full and q-pUCCD methods, together with the exact curve.
The OH distance is varied from the equilibrium value of $d=1.754$~\AA~to the final value of $d=2.393$~\AA~at fixed angle $\alpha= 104.51^{\circ}$. 
The RHF/STO-3G calculation produces 14 MOs among which 10 are occupied. 
The two $1s$ orbitals of the oxygen atom are then replaced by the corresponding frozen core potentials. 
Finally, the number of degrees of freedom is further reduced to 8 electrons in 9 orbitals (i.e., qubits) by applying tapering (see Section~\ref{sec:methods}) to the Hamiltonian in second quantization.
Note once more that all these operations do not affect the spectrum of the original Hamiltonian.

The projective CCSD method breaks down for distances $d>2$~\AA. In fact, we observe a non-physical barrier for re-binding from large distances that is not observed for the exact solution.
In contrast, as for the case of H$_4$, all quantum approaches show qualitatively the right behaviour, with the q-UCCSD method approaching very closely (within chemical accuracy) the exact curve. 
Note that all curves are shifted in order to match the initial point at $d=1.754$~\AA. The values of the shifts are reported in the legend and summarized in Tab.~\ref{tab:results}.
As expected, the limitations applied to the possible excitations (singlet and pair) induce a sizable up-shift of the curves as for H$_4$, the largest error obtained for q-pUCCD.  
The lower panels of Fig.~\ref{fig:H2O_all} report the relative energy errors with respect to the exact solution for all methods. 
Both the original q-UCCSD and the new q-UCCD0 Ans\"atze produce dissociation curves close to chemical accuracy over the entire dissociation profile.
The absolute energies improve substantially when the different approximations are applied together with OO as shown in the right panel of Fig.~\ref{fig:H2O_all}. 
In this case, the best results are obtained for the q-oo-UCCD0 Ansatz, which gives a maximum error of about $5.0$ mHa over the entire dissociation profile.
Most importantly, all approximations reproduce the correct qualitative monotonic behaviour with a similar NPE values (see Tab.~\ref{tab:results}), in contrast to the classical CCSD method, which is qualitatively wrong. 

Concerning the computational efficiency of the different  approaches, both q-UCCD0 and q-pUCCD require approximately half of the excitations needed for the full q-UCCSD Ansatz (see Tab.~\ref{tab:results} for a detailed account).
The optimization using the q-UCCD0-full Ansatz reduces the energy shift compared to q-UCCD0, at the cost of increasing the number of gates (Tab.~\ref{tab:results}).

The corresponding number of single- and two-qubit gates required for the implementation of these approaches follows the same tend, making the orbital-optimized pair Ansatz (q-oo-pUCCD) the favourable solution that maximizes the accuracy/cost ratio.
It is also important to stress that numerically the q-pUCCD and q-oo-pUCCD approaches need only one third of the total number of energy evaluations to achieve energy convergence using VQE with the SLSQP optimizer.

\subsubsection{N$_2$ Molecule}

The nitrogen molecule is one of the most severe test cases for single-reference electronic structure approaches due to the strong correlation character associated to the stretching of the triple bond.
Unlike the cases of the H$_4$ and the H$_2$O molecules, this system contains 6 active $p$ electrons contributing to the bond giving rise to multiple equally-weighted configurations at the dissociation limit. 

\begin{figure*}[ht]
\centering
\includegraphics[width=1.\linewidth]{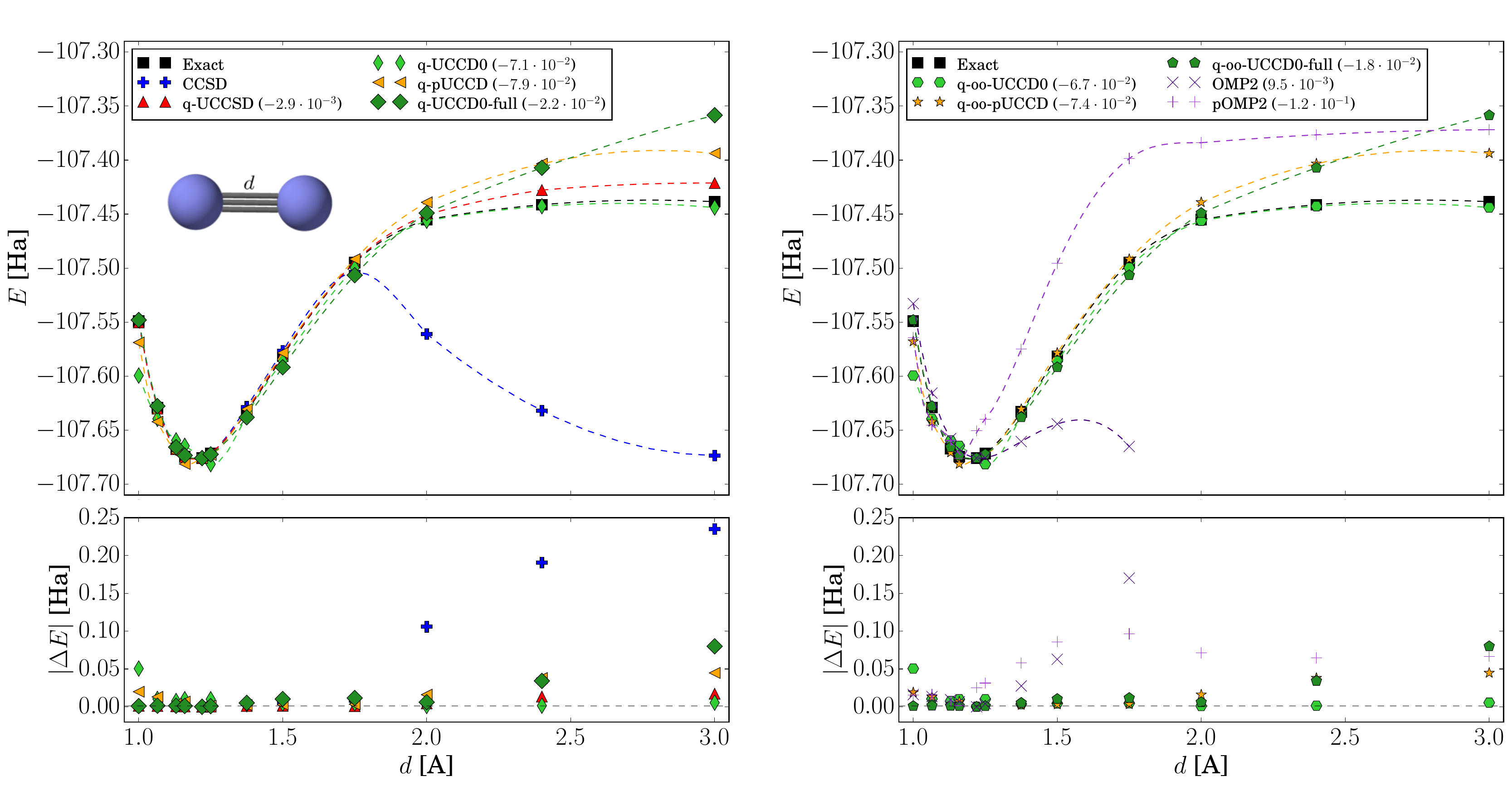}
\caption{
\it \textbf{Upper panels:}  Energy profiles of the \ce{N$_2$} molecule as a function of the bond length $d$ computed using different variants of the classical and quantum CC approaches. All profiles are shifted to match the exact curve at $d=1.2$~\AA. 
The shifts in Hartree are reported within parenthesis next to the acronyms labelling the different approaches (see also Table~\ref{tab:results}).
		\textbf{Lower panels:}
    Absolute energy differences with respect to the exact profile (obtained upon diagonalization of the Hamiltonian). The black dashed line corresponds to the chemical accuracy threshold at 1.6 mHa.}
\label{fig:N2_all}
\end{figure*}

Using the STO-3G basis set, we can describe the N$_2$ molecule with 14 electrons in 20 spin-orbitals.
Using the frozen-core approximation we further reduce the number of electrons and MOs by 4.
Finally, by applying tapering the problem size is reduced to 10 electrons in 12 orbitals, which maps to a quantum register with 12 qubits.

Fig.~\ref{fig:N2_all} shows the energy dissociation curves obtained upon stretching the N$_2$ bond length $d$ from 1 to 3~\AA.
The classical CCSD Ansatz fails to reproduce the correct dissociation profile for $d > 1.75$ \AA.
At these distances, the near-degenerate states acquire symmetry instabilities, which leads to a dramatic failure of the method. 
In fact, the energy, instead of asymptotically increasing, drops to a value comparable to the one of the equilibrium distance. 
The overall profile is therefore qualitatively wrong, showing a recombination barrier of the same size of the dissociation one. 
In order to overcome this problem, several `corrections' to the original CCSD Ansatz have been proposed in the literature such as the CCD0 approach~\cite{Bulik2015}. 
However, while curing at least qualitatively the dissociation profile, the discrepancy with the exact solution remains very large.

On the other hand, all the q-UCC can reproduce at least qualitatively the correct dissociation profile.
The q-UCCSD curve lies within 2 mHa in energy from the reference over the entire distance range (Fig.~\ref{fig:N2_all}, lower left-hand panel). 
Among the approximated q-UCCSD approaches, q-UCCD0 is the one that provides the best results in terms of average error (Fig.~\ref{fig:N2_all}), NPE and asymptotic behaviour (Tab.~\ref{tab:results}).  
However, the overall energy shifts applied in order to match all energy at the equilibrium bond length remain relatively large (between 20 and 80 mHa).
When applying OO the picture does not change substantially. q-oo-UCCD0 remains the more accurate wavefunction Ansatz and leads to only a minor improvement in  absolute energy compared to q-UCCD0.
Interestingly, as for the other molecules in our study set (namely H$_4$ and H$_2$O), the q-oo-pUCCD method produces a qualitatively correct profile with a maximal deviation from the reference of about 80 mHa and a NPE of 65 mHa.
For comparison, in the right panel of Fig.~\ref{fig:N2_all} we also report the results obtained using OMP2 and pOMP2.
Note that for distances close to 1 \AA \ (see Fig.~\ref{fig:N2_all}), the discrepancy between the q-UCCD0 and the exact results is relatively large, which directly reflects in the large NPE (60 mHa). 
On the other hand, when computed for distances larger than $1.2$ \AA, the q-UCCD0 Ansatz produces smaller NPE than q-UCCSD, indicating that this solution is to be preferred for stretched geometries (strongly correlated regime).

Also in this case, all OO approaches maintain or improve their efficiency (i.e. reducing the absolute error) compared to their reference (non-OO) approaches.
The origin of the increase of the NPE using q-oo-UCCD0 (observed for H$_4$ and N$_2$) and q-oo-UCCD0-full (in the case of H$_2$O) may be associated to the disruption of the balance among the double excitations introduced in the q-UCCD0 and q-UCCD0-full Ans\"atze with the purpose of improving the solution in the strongly correlated regime. In fact, OO modifies the nature of the orbitals and introduces (effective) single excitations (see Appendix B) making therefore the interpretation of the q-UCCD0 and q-UCCD0-full Ans\"atze less evident.
Further, the addition of new degrees of freedom i.e., the parameters used to define the orbital rotation matrix $\kappa$, introduces more complexity in the energy landscape (adding further local minima) hampering the numerical convergence towards the global energy minimum.
In conclusion, while the addition of the OO can clearly improve the total energy (by offering more degrees of freedom) the combination with the q-UCCD0 and q-UCCD0-full Ans\"atze may actually increase the NPE.

\subsubsection{Hubbard Model}
In this last application, we turn our attention to the investigation of a prototypical strongly correlated periodic  system described by the one dimensional Fermi-Hubbard Hamiltonian given in Eq.~\eqref{Hubbard}. 
This model was originally designed to study strong electronic interaction in narrow energy band materials~\cite{Hubbard1963}.

\begin{figure}[ht]
\centering
\includegraphics[width=1.\linewidth]{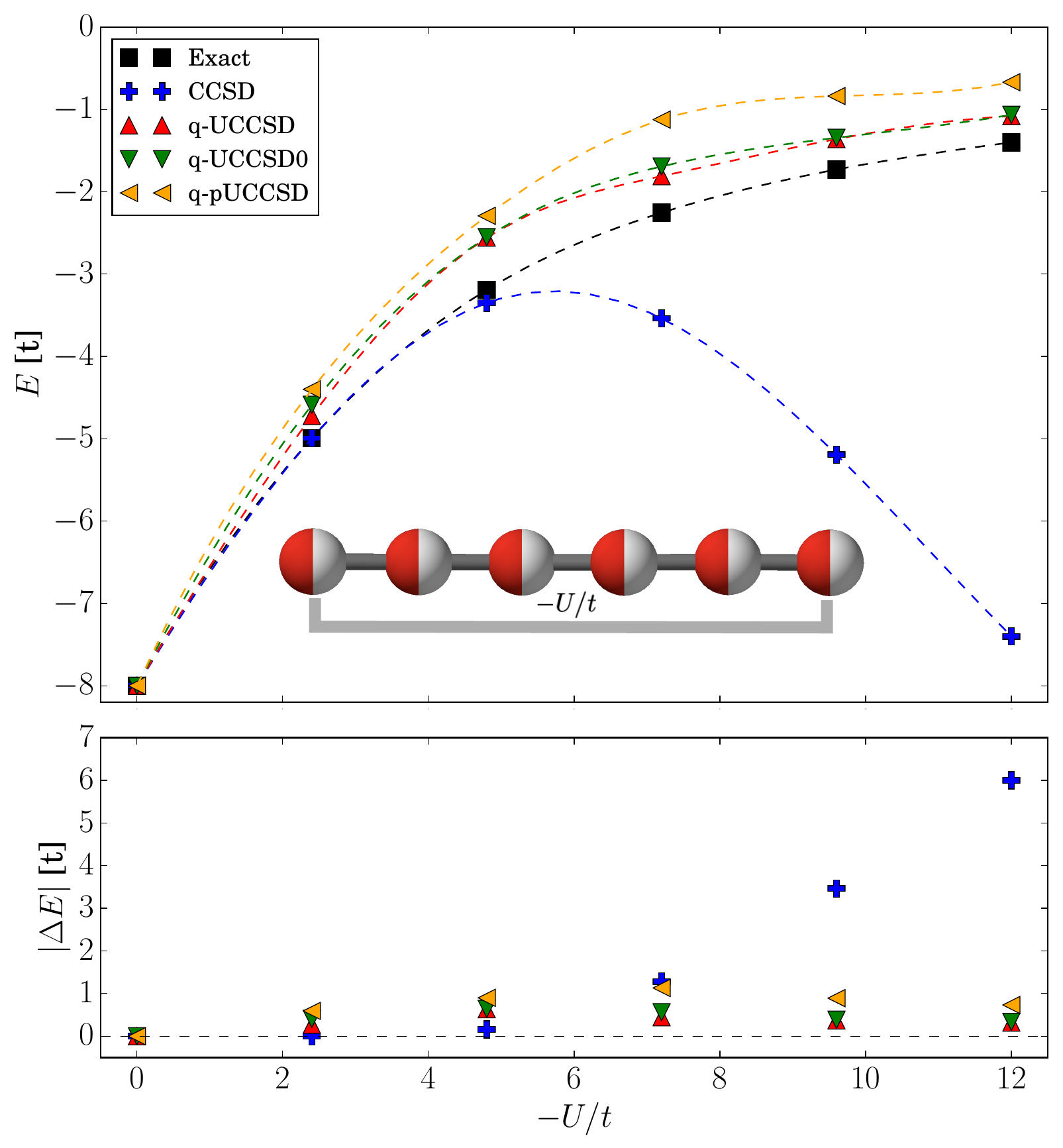}
\caption{
\it 
	\textbf{Upper panels:} Ground state energy (in units of electron hopping term $t$) for the periodic one-dimensional Fermi-Hubbard chain at half-filling as a function of the interaction energy $U$ at fixed $t=-1$ (see Eq.~\eqref{Hubbard}). The different lines correspond to different classical and quantum wavefunction Ans\"atze (see legend).
	The exact curve corresponds to the lowest eigenvalue of the corresponding Hamiltonian. 
	\textbf{Lower panels:}
    Absolute energy differences with respect to the exact diagonalisation profile. The black dashed line corresponds to zero energy.}
\label{fig:Hubbard_all}
\end{figure}

The inset of Fig.~\ref{fig:Hubbard_all} shows the 6-sites Fermi Hubbard model used in this study. 
The line connecting the first with the last site indicates the  periodicity of the system. 
Here we restrict our analysis to the half-filling (6 electrons in $N=6$ sites) scenario (see Eq.~\eqref{Hubbard}), which gives rise to a number of degenerate states close to the ground state solution. 
The implementation of the model in the qubit register is done by assigning two qubits for each site, one for spin-up and one for spin-down electrons. No qubit count reduction schemes such as tapering have been used in this case due to the absence of the required symmetries in the Hamiltonian.

Interestingly, by increasing the ratio $|U/t|$ In the Fermi Hubbard model we can control the transition towards a regime of strongly correlated electrons dominated by the two-body Coulomb repulsive term.
In Fig.~\ref{fig:Hubbard_all}, we monitor this transition sweeping the interaction parameter $U$ from 0 to 12 while keeping $t$ fix to the value $-1$.

In the following study, we investigate the variational ground state solution within the subspace with total spin $S=0$ (three spin-up and three spin-down electrons) of the full Hilbert space. This means that 
we do not allow for spin-flip excitations while the initial state is prepared in the $S=0$ subspace.
For the Hubbard model, the use of `generalized' excitations (see refs.~\cite{mcclean_theory_2016, Lee2018}) in the definition of the q-UCCSD Ansatz did not bring any significant advantage. Therefore, only `standard' single excitations are used for computational efficiency (i.e., to limit the number of variational parameters).
Note that the presence of a repulsive $U$ term in the Hamiltonian favours electronic configuration with only one electron per site. Therefore, the action of the single excitation operators is particularly important in the Hubbard model and they will be considered explicitly also in the case of the q-pUCCD Ansatz. 
As explained in Section~\ref{subsec:subsection_ooCC_appro} and Appendix B, this will replace the use of OO.

As expected, the CCSD Ansatz breaks down as soon as the multi-reference character starts dominating the ground state wavefunction at $-U/t > 4$. 
On the other hand, all quantum models including the approximated ones (i.e., q-UCCSD0 and  q-pUCCSD) can reproduce, at least qualitatively, the correct asymptotic behaviour, in agreement with the exact solution. 
Also in this case, the full q-UCCSD method performs better than q-UCCSD0 and q-pUCCSD, as confirmed by the error plots in lower panel of Fig.~\ref{fig:Hubbard_all}. 

Contrary to the classical projective CC method where the modified CCD0 and pCCD approximations were introduced to cure failures of original CCSD Ansatz, in the quantum case the approximated q-UCCSD0 and q-pUCCSD methods do not improve the accuracy of the solution.
However, while less accurate, they provide a qualitatively correct description of the system at a much lower computational cost. 
As shown in Tab.~\ref{tab:results}, the number of total iterations and energy evaluations are systematically reduced when going from the original q-UCCSD approach to the approximated q-UCCSD0 and q-pUCCSD methods.

\section{\label{sec:conclusion}Conclusion}

Coupled Cluster (CC) is a single-reference post-Hartree-Fock wavefunction approach capable of reproducing accurate electronic structure properties within chemical accuracy for many molecular systems. 
However, in its projective formulation, CC is known to fail when dealing with strongly correlated systems dominated by static correlation. 
In this case, multi-reference extensions of the CC Ansatz are required (MRCC) with a corresponding increase of the computational costs.

In this article, we showed that the variants of quantum unitary CC method (named q-UCC), obtained from the transposition of the classical UCC Ansatz, can produce qualitatively correct energy profiles also in the strongly correlated regime.
In particular, we focused our study on the investigation of the properties of the q-UCCSD Ansatz for which the CC expansion is truncated to single and double excitations.
The quantum algorithm is obtained by encoding the UCCSD Ansatz as series of parametrized qubit operations in a quantum register using a trotterization of the cluster operator 
$
e^{\hat{T}(\vec{\theta})-\hat{T}^{\dagger}(\vec{\theta})} 
$ (see Eq.~\eqref{eq:UCCSD_trotter}), where $\hat T$ is the excitation operator.  
The main reason for the success of this algorithm lies in its variational nature, which enables to recover from several shortcomings of the projective CCSD method, including the capability to deal with strongly correlated systems.

To demonstrate the quality of the q-UCCSD predictions, we computed the energy profiles for a series of molecular systems along reaction paths that lead to the stretching of one or more bonds. 
In this way, we were able to monitor the accuracy of the q-UCCSD Ansatz in both, the equilibrium (single reference) regime and the dissociative (multi-reference and strongly correlated) limit.
For all investigated systems, the molecules \ce{H$_4$}, \ce{H$_2$O},  \ce{N$_2$}, and the one-dimensional Fermi-Hubbard chain, we obtained accurate results that outperform the classical projective CCSD method.
In particular, all our simulations showed the correct qualitative dissociation profiles and in most cases (with the only exception of a few points in the dissociation of \ce{N$_2$}) the energy differences with the reference curves were within chemical accuracy (i.e. $< 1.6$ mHa) over the entire dissociation profile.

Moreover, we extended our investigation to the analyses of approximated q-UCCSD Ans\"atze for which the number of possible one- and two-electron excitations has been limited in order to reduce the total number of gate operations and the corresponding variational parameters to optimize. 
This is an important prerequisite in view of the possible future implementation of these approaches in near-term quantum computers.
To this end, we explored two main reduction schemes, which limit the nature of the double excitations to a subset of all possible ones.
The singlet and the pair q-UCCD approaches (named q-UCCD0 and q-pUCCD, respectively) and their corresponding orbital-optimized versions (that were already proposed as alternatives to projective CCSD) showed interesting results for all tested systems, in qualitative agreement with the corresponding reference curves.
In particular, orbital optimization restores a significant fraction of the correct absolute value of the total energy at the chosen reference geometries while reproducing, at least qualitatively, the correct energy profiles within 10-20 mHa accuracy from the reference calculations. 
More importantly for the future of near-term quantum computing, these results are achieved using a fraction (e.g. between $2/3$ for H$_4$ and $1/4$ for H$_2$O in the case of the q-oo-pUCCD Ansatz) of the number of two-qubit gates required for the original q-UCCSD approach.
Further investigations are needed to enable the efficient implementation of q-UCCSD approaches in near-term quantum devices~\cite{bauman2019downfolding, grimsley2019adaptive, grimsley2019trotterized}.

In conclusion, we demonstrated the potential of the q-UCCSD Ansatz (for VQE) and the advantages with respect to the classical equivalent (projective CCSD), especially in the description of systems with strong electronic correlation.
Finally, the variational nature of the VQE and ooVQE algorithms in combination with the approximations of the q-UCCSD Ansatz discussed in this work can open up new possibilities for the solution of electronic structure problems (also in the strongly correlated regime) of medium to large molecular systems using shallow circuits in near-term noisy quantum computers.

\section{\label{sec:acknowledgements}Acknowledgements}
The authors acknowledge the financial support from the Swiss National Science Foundation (SNF) through the grant No. 200021-179312.
We also thank Matthieu Gilles Mottet, Almudena Carrera Vazquez, Luca Erhart, Anton Robert, John West, Guglielmo Mazzola and Manfred Sigrist for stimulating discussions.
IBM, IBM Q, Qiskit are trademarks of International Business Machines Corporation, registered in many jurisdictions worldwide. Other product or service names may be trademarks or service marks of IBM or other companies.

\section*{\label{sec:appendixA}Appendix A}
\subsection*{\label{sec:appendixA_ints} Molecular Integrals}
The coefficients of the Hamiltonian in Eq.~\ref{eq:H_sec_quant_expl} are given by one-electron integrals defined in physics notation on molecular orbital basis as
\begin{equation}
 \langle r |\hat h| s \rangle = \int dr_1 \, \phi_r^*(r_1) \left(-\frac{1}{2} \nabla^2_{r_1} - \sum^M_{I=1} \frac{Z_I}{R_{1I}}\right) \phi_s(r_1) 
\end{equation}
and similarly for the two-electron terms given by
\begin{equation}
\langle r s |\hat g| t u \rangle = \int d r_1 dr_2 \, \phi_r^*(r_1) \phi_s^*(r_2) \frac{1}{r_{12}} \phi_t(r_1) \phi_u(r_2) \, . 
\end{equation}
where $Z_I$ is the nuclear charge of atom $I$, $R_{1I}=|R_I-r_1|$, and $r_{12}=|r_1-r_2|$.

\section*{\label{sec:appendixB}Appendix B}
\subsection*{\label{sec:appendixB_orb_rot}Orbital rotations}
Orbital rotations can be defined by means of a unitary transformation $R$ acting on the orbitals in first quantization
\begin{equation}
\phi'_{r}(r)=\sum_{s} R_{rs} \, \phi_s(r)
\end{equation}
where the $R$ can be written in terms of an antihermitian matrix $\kappa$
\begin{equation}
R=\exp{(- \kappa)}
\label{eq:orbital_rotation_matrix}
\end{equation}
with $\kappa^{\dagger}=-\kappa$.
Therefore, the orbital rotation can be applied to one-/two-electron integrals by means of a basis change 
\begin{align}
  \langle r |\tilde{\hat{h}}| s \rangle &= \sum_{ab}C^{\alpha, *}_{ar} C^{\alpha}_{bs} \langle a |\hat{h}| b \rangle  \label{eq:h_rotated} \\
  \langle p q |\tilde{\hat{g}}| r s \rangle &= \sum_{abcd}C^{\alpha, *}_{ap} C^{\alpha, *}_{bq} C^{\alpha}_{cr} C^{\alpha}_{ds} \langle a b |\hat{g}| c d \rangle \label{eq:g_rotated}
\end{align}
where $a,b,c,d$ label the AOs and $p,q,r,s$ the MOs.
By acting on AO-to-MO coefficient matrices $C^{\alpha}$ and $C^{\beta}$ with matrix $R$, we obtain in the RHF case
\begin{align}
C^{\alpha}&={C}_{\text{RHF}}^{\alpha} \exp{(-\kappa_{\alpha})} \label{eq:Calpha} \\
C^{\beta}&={C}_{\text{RHF}}^{\beta} \exp{(-\kappa_{\beta})}. \label{eq:Cbeta}
\end{align}
This transformation preserves orthonormality due to unitarity of $R$.
In addition, the RHF orbtials are real and the spin-restriction forces $\kappa = \kappa_{\alpha} = \kappa_{\beta}$ which implies $C^{\alpha}=C^{\beta}$.

The same transformation can be equivalently applied to the creation and annihilation operators (for second quantization, see Sec. 3.2 of ref.~\cite{Helgaker2014})
\begin{align}
    \tilde{\hat{a}}_r &=\exp{(-\hat\kappa)} \hat{a}_r \exp{(\hat\kappa)} = \sum_{s} R^*_{sr} {\hat{a}}_s \\
    \tilde{\hat{a}}^{\dagger}_r &=\exp{(-\hat\kappa)} \hat{a}^{\dagger}_r \exp{(\hat\kappa)} = \sum_{s} R_{sr} {\hat{a}}^\dagger_s \label{eq:rot_ops}
\end{align}
where we introduced the anti-Hermitian operator
\begin{equation}
\hat\kappa = \sum_{rs} \kappa_{rs} \hat{a}^{\dagger}_r \hat{a}_s \label{eq:kappa_operator}
\end{equation}
We can therefore write the Hamiltonian on the rotated orbital basis as
\begin{align}
{\hat{H}'} =&\sum_{rs} \langle r |\tilde{\hat{h}}| s \rangle \,  {\hat{a}}^{\dagger}_r {\hat{a}}_s  \\ &+
\frac{1}{2}\sum_{rstu}\langle r s |\tilde{\hat{g}}| t u \rangle \,   {\hat{a}}^{\dagger}_r {\hat{a}}^{\dagger}_s {\hat{a}}_u {\hat{a}}_t + E_{NN} \notag
\label{eq:H_sec_quant_rotated}
\end{align}

We arrive at the conclusion that applying the rotation to the orbitals in first quantization is equivalent to applying the following transformation to the 
Hamiltonian in second quantization 
\begin{equation}\label{eq:H_orbital_rotated}
\hat H \rightarrow \hat H'
= \exp{(-\hat\kappa)} \hat{H} \exp{(\hat\kappa)}
\end{equation}
which can be also associated to
\begin{equation}
\hat H \rightarrow \hat H'= \exp{(-(\hat{T}_1-\hat{T}_1^{\dagger}))} {\hat{H}} \exp{(\hat{T}_1-\hat{T}_1^\dagger)}
\end{equation}
for a generalized single excitation $\hat{T}_1$ operator (all possible excitations are allowed).  
By construction, both Hamiltonians $\hat{H}$ and $\hat{H}'$ share the same spectrum. 
Therefore, thanks to the optimization of orbitals through rotations we aim for a maximal overlap of the exact solution with the support specified by the selected wavefunction Ansatz.

\section*{Appendix C}

\subsection*{Effect of the Orbital Optimization on the MOs}

As an example of the implementation of the OO approach described in Section~\ref{sec:methods}, we take the H$_4$ molecule at the geometry corresponding to $R=1.738$~\AA \ and $\beta = 85^{\circ}$.

\begin{figure*}[ht]
\centering
\includegraphics[width=0.85\linewidth]{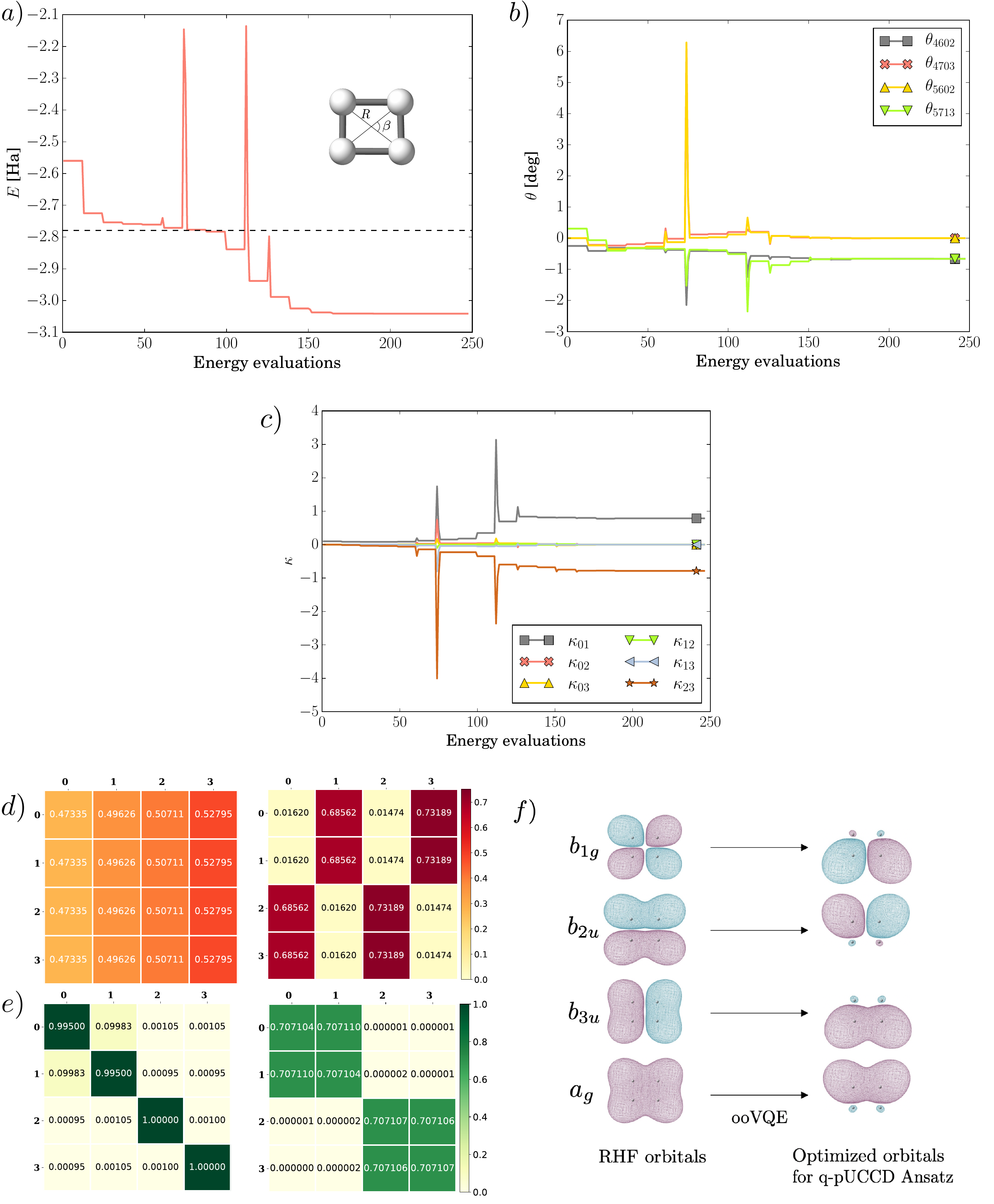}
\caption{\it Computation of the ground state of H$_4$ molecule at the geometry corresponding to $R=1.738$~\AA, $\beta = 85^{\circ}$ using the q-(oo-)pUCCD Ansatz and ooVQE algorithm.
\textbf{a)} Evolution of the energy during the execution of ooVQE algorithm. 
The dashed line corresponds to the optimized energy obtained without OO. 
\it \textbf{b)} Evolution of the elements of $\vec\theta$ vector associated with double excitations during optimization. 
\it \textbf{c)} Evolution of the elements of $\vec\kappa$ vector associated with rotation matrix for the MO during optimization.
\it \textbf{d)} Absolute values of the elements of the rotation matrix $C$ (Eqs.~\eqref{eq:Calpha} and~\eqref{eq:Cbeta}) before ($C=C_{\text{RHF}}$, left) and after convergence ($C=C_{\text{RHF}} \, e^{-\kappa}$, right).
\it \textbf{e)} Same as in d) but for the absolute value of the elements of the antihermitian matrix $\kappa$ (see Appendix B).
\it \textbf{f)} Representation of the initial and optimized MOs obtained after applying the ooVQE algorithm.}
\label{fig:H4_orb_rot}
\end{figure*}

Fig.~\ref{fig:H4_orb_rot} reports the time evolution of the  optimization parameters ($\vec{\theta}, \vec{\kappa}$) using ooVQE algorithm. We observe that while the energy of the system is converging (panel a)) both the CC variational parameters (panel b)) and the orbital rotation matrix elements (panel c)) are stabilizing on plateau values. The steps and spikes in the different profiles are induced by re-settings of the classical optimization algorithm.

The initial and final values of the elements of the rotation matrix $C$ (Eqs.~\eqref{eq:Calpha} and~\eqref{eq:Cbeta}) are given in panel d) from left to right.
The equivalent plot for the absolute value of the elements of the antihermitian matrix $\kappa$ are reported in Figure e). We observe that the action of the OO procedure changes substantially the symmetry of matrices introducing a $2\times2$ block structure.

\section*{\label{sec:appendixD}Appendix D}

\subsection*{Effect of Tapering Off Qubits on Required Quantum Resources}

In Tab.~\ref{tab:results_tapering} we report the number of qubits and quantum gates used when employing tapering~\cite{Bravyi2017a} for the \ce{H_4}, \ce{H_2O} and the \ce{N_2} molecules.

\begin{table*}[ht]
\centering
\caption[]
{
\textit{Number of qubits and quantum gates for the different q-UCC Ans\"atze after tapering off qubits. 
Description of the abbreviations: 
Hub.: Hubbard;
No.: number; 
SQG: single qubit gates, u1,u2,u3 corresponding to one-,two-, and three-parameters one-qubit gates (details are presented in ref.~\cite{qiskit2019});
TQG: two-qubit gates, CNOT gates.
}}
\label{tab:results_tapering}
\scriptsize
\begin{tabular}{c c c c c c}
\hline \hline
 &    & q-UCCSD & q-UCCD0 & q-pUCCD & q-UCCD0-full  \\
\hline
 H$_4$      & No.\ SQG    & 	64/204/185	&		64/236/189	&	24/92/77	 	&	 	112/440/345  \\
   (5 qubits) & No.\ TQG           &    	364	&	 	404	&	156 	&	 728     \\ 
\hline
 H$_2$O        & No.\ SQG       &   	376/2344/1339	&	 	240/1068/916	&	80/816/310 	&	 496/3824/1843     \\
(9 qubits) & No.\ TQG           &    	11932	&	 3000	&	1168	 	&	  6048      \\ 

\hline
N$_2$      & No.\ SQG     &       	780/6432/3057	&	720/6272/3910	&   120/1456/471	 	&	 	1800/16400/10246      \\
(12 qubits)     & No.\ TQG              &    	9200	&		11792	&	2032	    &     30160		 \\ 
\hline \hline
\end{tabular}
\end{table*}

\section*{\label{sec:appendixE}Appendix E}

\subsection*{Example of Quantum Singlet UCCD (q-UCCD0) and Quantum Singlet Full UCCD (q-UCCD0-full)}

To clarify the difference between the q-UCCD0 and q-UCCD0-full Ans\"atze, we consider the simplest non-trivial case of 2 electrons in 6 MOs (i.e. 3 spatial orbitals with $0$th being doubly occupied; $1$st and $2$nd being virtual). 
In this setting, the singlet double excitation operator (Eq.~\eqref{eq:t1_singlet}) becomes
\begin{align}
\label{eq:t2_uccd0full1}
\hat{T}_2^{0}(\vec{\alpha})&=  \, 2\alpha_{1100} \, ( \hat{a}^{\dagger}_{1,\downarrow}\hat{a}^{\dagger}_{1,\uparrow}\hat{a}_{0,\downarrow}\hat{a}_{0,\uparrow}) \\ &\quad +  (\alpha_{1200}+\alpha_{2100})\,\notag \\ & \quad \times(\hat{a}^{\dagger}_{1,\downarrow}\hat{a}^{\dagger}_{2,\uparrow}\hat{a}_{0,\downarrow}\hat{a}_{0,\uparrow}+\hat{a}^{\dagger}_{2,\downarrow}\hat{a}^{\dagger}_{1,\uparrow}\hat{a}_{0,\downarrow}\hat{a}_{0,\uparrow}) \notag \\ & \quad +  2\alpha_{2200} \, ( \hat{a}^{\dagger}_{2,\downarrow}\hat{a}^{\dagger}_{2,\uparrow}\hat{a}_{0,\downarrow}\hat{a}_{0,\uparrow}). \notag
\end{align}
The symmetries of the $\hat{T}^{0}_{2}$ tensor can be exhibited by considering independent amplitudes for each excitation (e.g. $\theta_{{1}_\downarrow {2}_\uparrow {0}_\downarrow {0}_\uparrow}$ corresponding to $\hat{a}^{\dagger}_{1,\downarrow}\hat{a}^{\dagger}_{2,\uparrow}\hat{a}_{0,\downarrow}\hat{a}_{0,\uparrow}$) in Eq.~\eqref{eq:t2_uccd0full1} which reads
\begin{align}
\label{eq:t2_uccd0full2}
\hat{T}_2^{0}(\vec{\theta}) &= \, 2\theta_{{1}_\downarrow {1}_\uparrow {0}_\downarrow {0}_\uparrow} \, ( \hat{a}^{\dagger}_{1,\downarrow}\hat{a}^{\dagger}_{1,\uparrow}\hat{a}_{0,\downarrow}\hat{a}_{0,\uparrow}) \notag \\ &\quad + \, \theta_{{1}_\downarrow {2}_\uparrow {0}_\downarrow 0\uparrow} \hat{a}^{\dagger}_{1,\downarrow}\hat{a}^{\dagger}_{2,\uparrow}\hat{a}_{0,\downarrow}\hat{a}_{0,\uparrow} \\ & \quad +\theta_{{2}_\downarrow {1}_\uparrow {0}_\downarrow {0}_\uparrow}\hat{a}^{\dagger}_{2,\downarrow}\hat{a}^{\dagger}_{1,\uparrow}\hat{a}_{0,\downarrow}\hat{a}_{0,\uparrow} \notag \\ & \quad +  2 \theta_{{2}_\downarrow {2}_\uparrow {0}_\downarrow {0}_\uparrow} \, ( \hat{a}^{\dagger}_{2,\downarrow}\hat{a}^{\dagger}_{2,\uparrow}\hat{a}_{0,\downarrow}\hat{a}_{0,\uparrow}). \notag
\end{align}
This allows us to identify the coefficients as $(\alpha_{1200}+\alpha_{2100})/{2} = \theta_{1_{\downarrow} 2_{\uparrow} 0_{\downarrow} 0_{\uparrow}} = \theta_{2_{\downarrow} 1_{\uparrow} 0_{\downarrow} 0_{\uparrow}}$. 
Using this relation, we can simplify Eq.~\eqref{eq:t2_uccd0full2} by absorbing the scalars with the amplitudes. 
As a consequence, Eq.~\eqref{eq:t2_uccd0full2} becomes
\begin{align}
\label{eq:t2_uccd0full3}
\hat{T}_2^{0}(\vec{\theta}) &= \, \theta_{{1}_\downarrow {1}_\uparrow {0}_\downarrow {0}_\uparrow} \,  \hat{a}^{\dagger}_{1,\downarrow}\hat{a}^{\dagger}_{1,\uparrow}\hat{a}_{0,\downarrow}\hat{a}_{0,\uparrow} \\ &\quad + \, \theta_{{1}_\downarrow {2}_\uparrow {0}_\downarrow 0\uparrow}(\hat{a}^{\dagger}_{1,\downarrow}\hat{a}^{\dagger}_{2,\uparrow}\hat{a}_{0,\downarrow}\hat{a}_{0,\uparrow} +\hat{a}^{\dagger}_{2,\downarrow}\hat{a}^{\dagger}_{1,\uparrow}\hat{a}_{0,\downarrow}\hat{a}_{0,\uparrow}) \notag \\ & \quad +  \theta_{{2}_\downarrow {2}_\uparrow {0}_\downarrow {0}_\uparrow} \, \hat{a}^{\dagger}_{2,\downarrow}\hat{a}^{\dagger}_{2,\uparrow}\hat{a}_{0,\downarrow}\hat{a}_{0,\uparrow} \notag
\end{align}
Note that in this way we achieved a reduction of the number of amplitudes from 4 (Eq.~\eqref{eq:t2_uccd0full2}) to 3 (Eq.~\eqref{eq:t2_uccd0full3}).

Furthermore, as the MOs involved in the double excitations $\hat{a}^{\dagger}_{1,\downarrow}\hat{a}^{\dagger}_{2,\uparrow}\hat{a}_{0,\downarrow}\hat{a}_{0,\uparrow}$ and $\hat{a}^{\dagger}_{2,\downarrow}\hat{a}^{\dagger}_{1,\uparrow}\hat{a}_{0,\downarrow}\hat{a}_{0,\uparrow}$ are identical within the RHF framework, we can further reduce the number of excitations (and therefore quantum gates) by considering explicitly only one of the two, e.g., $\hat{a}^{\dagger}_{2,\downarrow}\hat{a}^{\dagger}_{1,\uparrow}\hat{a}_{0,\downarrow}\hat{a}_{0,\uparrow}$. 
Consequently, we can defined the reduced $\hat{T}_2^{0}$ operator as
\begin{align}
\label{eq:t2_uccd0}
\hat{T}_2^{0, \Omega}(\vec{\theta}) &= \, \theta_{{1}_\downarrow {1}_\uparrow {0}_\downarrow {0}_\uparrow} \,  \hat{a}^{\dagger}_{1,\downarrow}\hat{a}^{\dagger}_{1,\uparrow}\hat{a}_{0,\downarrow}\hat{a}_{0,\uparrow} \notag \\ &\quad + \, \theta_{{1}_\downarrow {2}_\uparrow {0}_\downarrow 0\uparrow}\hat{a}^{\dagger}_{1,\downarrow}\hat{a}^{\dagger}_{2,\uparrow}\hat{a}_{0,\downarrow}\hat{a}_{0,\uparrow} \\ & \quad +  \theta_{{2}_\downarrow {2}_\uparrow {0}_\downarrow {0}_\uparrow} \, \hat{a}^{\dagger}_{2,\downarrow}\hat{a}^{\dagger}_{2,\uparrow}\hat{a}_{0,\downarrow}\hat{a}_{0,\uparrow}. \notag
\end{align}
Note that due to the four-fold symmetry in $\hat{T}^{0}_2$ tensor (Eq.~\eqref{eq:anlges_sym0}), we can at most eliminate 3 excitations out of 4. 
In order to provide physical insight into this procedure, we also consider the energy contribution associated to each individual excitation $\hat{t}\in\{ \hat{a}^{\dagger}_{m, \sigma}\hat{a}^{\dagger}_{n, \tau}\hat{a}_{i, \sigma}\hat{a}_{j, \tau} \}$ by minimizing the total energy $E(\theta)=\langle \Phi_0 | e^{\theta (\hat{t}_2^{\dagger} - \hat{t}_2)} \hat{H} e^{\theta (\hat{t}_2 - \hat{t}_2^{\dagger} )} |\Phi_0 \rangle$.   
In Table~\ref{tab:excitations_energy_h2}, we report the results in case of H$_2$ molecule in 6-31G basis set. 
In general, the optimized energies for the different subsets are different. 
However, within each subset, each excitation contributes identically to the energy, which motivates our choice of keeping only one of the two (with adapted weight).

Finally, the q-UCCD0 or q-UCCD0-full wavefunction Ans\"atze can be constructed by substitution the cluster operators $\hat{T}_2^{0, \Omega}$ (Eq.~\eqref{eq:t2_uccd0}) respectively $\hat{T}_2^{0}$ (Eq.~\eqref{eq:t2_uccd0full3}) into Eq.~\eqref{eq:UCCSD_trotter}.

\begin{table}[ht]
\centering
\caption[]
{\textit{Double excitations operators entering the definition of $\hat{T}_2^{0}$ for the case of the H$_2$ molecule
in 6-31G basis set ($N_{MO}=8$). 
The excitations are grouped into 2 types: paired and singlet. 
For each excitation, we consider a unique q-UCCD Ansatz (e.g. $e^{\theta_{{1}_\downarrow {1}_\uparrow {0}_\downarrow {0}_\uparrow} ( \hat{a}^{\dagger}_{1,\downarrow}\hat{a}^{\dagger}_{1,\uparrow}\hat{a}_{0,\downarrow}\hat{a}_{0,\uparrow} -  \hat{a}_{0,\uparrow}^{\dagger}\hat{a}_{0,\downarrow}^{\dagger}\hat{a}_{1,\uparrow}\hat{a}_{1,\downarrow} )} |\Phi_0 \rangle$ for the first row of this Table) and report the minimized energy corresponding to a bond distance of $0.546$ \AA \, using the VQE algorithm. The paired subset is constituted by double excitations that involve only 2 spatial orbitals (see Fig.~\ref{fig:advanced_ucc}). 
The remaining double excitations can be organized in groups of two excitations according to the symmetries given in Eq.~\eqref{eq:anlges_sym0}. Each pair produce exactly the same energy.}}
\label{tab:excitations_energy_h2}
\small
\begin{tabular}{c c c}
\hline \hline
Subset type & Excitation & Energy [Ha]  \\ \hline
Paired  & $\hat{a}^{\dagger}_{1,\downarrow}\hat{a}^{\dagger}_{1,\uparrow}\hat{a}_{0,\downarrow}\hat{a}_{0,\uparrow}$ & $-1.090$ \\ \hline
Paired  & $\hat{a}^{\dagger}_{2,\downarrow}\hat{a}^{\dagger}_{2,\uparrow}\hat{a}_{0,\downarrow}\hat{a}_{0,\uparrow}$  & $-1.094$   \\ \hline
Paired  & $\hat{a}^{\dagger}_{3,\downarrow}\hat{a}^{\dagger}_{3,\uparrow}\hat{a}_{0,\downarrow}\hat{a}_{0,\uparrow}$  & $-1.095$   \\ \hline
Singlet  & $\hat{a}^{\dagger}_{2,\downarrow}\hat{a}^{\dagger}_{1,\uparrow}\hat{a}_{0,\downarrow}\hat{a}_{0,\uparrow}$  & $-1.088$   \\ 
 & $\hat{a}^{\dagger}_{1,\downarrow}\hat{a}^{\dagger}_{2,\uparrow}\hat{a}_{0,\downarrow}\hat{a}_{0,\uparrow}$ &  $-1.088$  \\ \hline
Singlet  & $\hat{a}^{\dagger}_{3,\downarrow}\hat{a}^{\dagger}_{1,\uparrow}\hat{a}_{0,\downarrow}\hat{a}_{0,\uparrow}$  & $-1.090$   \\ 
 & $\hat{a}^{\dagger}_{1,\downarrow}\hat{a}^{\dagger}_{3,\uparrow}\hat{a}_{0,\downarrow}\hat{a}_{0,\uparrow}$ &  $-1.090$  \\ \hline
Singlet  & $\hat{a}^{\dagger}_{2,\downarrow}\hat{a}^{\dagger}_{3,\uparrow}\hat{a}_{0,\downarrow}\hat{a}_{0,\uparrow}$  & $-1.088$   \\ 
 & $\hat{a}^{\dagger}_{3,\downarrow}\hat{a}^{\dagger}_{2,\uparrow}\hat{a}_{0,\downarrow}\hat{a}_{0,\uparrow}$ &  $-1.088$  \\
\hline \hline 
\end{tabular}
\end{table}

\section*{\label{sec:appendixF}Appendix F}

\subsection*{Tapering Off Qubits in a Fermionic Hamiltonian}

In this section, we give a summary of the tapering off qubits procedure described in the original work by S. Bravyi et al.~\cite{Bravyi2017a}

We consider a generic fermionic $M$-qubit Hamiltonian obtained after applying the Jordan-Wigner transformation,
\begin{equation}
\label{eq:qubitHam1}
\hat{H}_{q} = \sum_{i} h_i \boldsymbol{\hat\sigma}_i
\end{equation}
where $\boldsymbol{\hat\sigma}_i$ are $M$-qubit Pauli strings belonging to the set
\begin{equation}
 \mathcal{P}_M=\pm \{ \hat{I},\hat{\sigma}^x, \hat{\sigma}^y,\hat{\sigma}^z\}^{\otimes M} \, .
\end{equation}
The $M$-qubit Clifford group, $\mathcal{C}$ is defined as the set of unitary operators, $\hat{U}$ such that
\begin{equation}
 \hat{U} \boldsymbol{\hat\sigma} \hat{U}^{\dagger} \in \mathcal{P}_M
\end{equation}
for all $\boldsymbol{\hat\sigma} \in \mathcal{P}_M$.

We define a symmetry group $\mathcal{S} \in \mathcal{P}_M$ of a Hamiltonian $\hat{H}_{q}$ as the set of operations (excluding $-\hat{I}$) that commute with each Pauli string in $\hat{H}_{q}$.
It was shown~\cite{gottesman1997stabilizer} that for any abelian group $\mathcal{S} \subseteq \mathcal{P}_M$ with $-\hat{I} \notin \mathcal{S}$  there is a set of independent generators $\boldsymbol{\hat\tau}_1, \dots, \boldsymbol{\hat\tau}_k$ such that
\begin{equation}
 \boldsymbol{\hat\tau}_i = \hat{U} \hat{\sigma}_i^x \hat{U}^{\dagger}, \quad \quad i=1, \dots, k
\end{equation}
for some Clifford unitary operator $\hat{U} \in \mathcal{P}_M$.
Note that we arrange the qubits in such a way that the symmetry operations apply to the first $k$ qubits.

The transformed Hamiltonian becomes
\begin{equation}
\label{eq:qubitHam2}
\hat{H}'_{q} = \hat{U}^{\dagger} \hat{H}_{q} \hat{U} = \sum_{j} h_j \boldsymbol{\hat\eta}_j
\end{equation}
with $\boldsymbol{\hat\eta}_j= \hat{U}^{\dagger} \boldsymbol{\hat\sigma}_j \hat{U} \in \mathcal{P}_M$.

By construction, the two Hamiltonians $\hat{H}_{q}$ and $\hat{H}'_{q}$ are isospectral. 
However, by exploiting the commutation relations among the Pauli strings $\boldsymbol{\hat\sigma}_i$ and $\boldsymbol{\hat\eta}_j$, it is possible to replace operators (more specifically $\hat{I}$ and $\hat\sigma^x$ operators) within the first $k$ positions of the $\boldsymbol{\hat\eta}$ Pauli string with their corresponding eigenvalues achieving an effective reduction in the number of qubits. 
Following closely ref.~\cite{Bravyi2017a}, since $[\boldsymbol{\hat\tau}_i,\boldsymbol{\hat\sigma}_j]=0$ for all $i,j$ one has $[{\hat\sigma}^x_i,\boldsymbol{\hat\eta}_j]=0$ for all $i,j$. 
This implies that all terms in  $\hat{H}'_{q} $ must commute with the $\hat{\sigma}^x_i$ operators (with $i\leq k$) in each Pauli string $\boldsymbol{\hat\eta}_j$, or equivalently that the first $k$ terms of each Pauli string $\boldsymbol{\hat\eta}_j$ that appear in $\hat{H}'_{q} $ are $\in \{\hat{I}, \hat{\sigma}^x\}$. 
When looking for the ground energy of $\hat{H}'_{q}$ using a variational approach such as VQE one can therefore replace in the Pauli strings the $\hat{\sigma}^x_1,\dots,\hat{\sigma}^x_k$ operators by their eigenvalues $\pm 1$ and remove the first $k$ qubits of the qubit register.

For more details about the theory and the implementation of this procedure, the interested reader is invited to read the original paper of S. Bravyi et al~\cite{Bravyi2017a}.

\bibliographystyle{achemso}
\bibliography{libraryedited}

\end{document}